\documentclass[referee]{jfm}
\usepackage{natbib}
\usepackage{graphicx}

\title[Dynamics of sheared suspensions]{Deterministic and stochastic behaviour of 
non-Brownian spheres in sheared suspensions}

\author[G. Drazer, J. Koplik, B. Khusid and A. Acrivos]{G\ls E\ls R\ls M\ls A\ls N\ns 
D\ls R\ls A\ls Z\ls E\ls R$^1$,\ns J\ls O\ls E\ls L\ns K\ls O\ls 
P\ls L\ls I\ls K$^1$, \ns B\ls O\ls R\ls I\ls S\ns K\ls H\ls U\ls S\ls I\ls D$^2$\ns 
\and A\ls N\ls D\ls R\ls E\ls A\ls S\ns A\ls C\ls R\ls I\ls V\ls O\ls S$^1$}

\affiliation{$^1$The Levich Institute, T-1M, The City College of the
City University of New York,\\ New York, NY 10031, USA \\[\affilskip]
$^2$Department of Mechanical Engineering, New Jersey Institute of Technology, 
\\ University Heights, Newark, New Jersey 07102}
\pubyear{1993}
\volume{}
\pagerange{}
\date{\today}
\begin{document}

\maketitle

\begin{abstract}
The dynamics of macroscopically homogeneous sheared suspensions of 
neutrally buoyant, non-Brownian spheres is investigated in the limit
of vanishingly small Reynolds numbers using Stokesian dynamics.
We show that the complex dynamics of sheared suspensions can be characterized 
as a chaotic motion in phase space and determine the dependence of the largest 
Lyapunov exponent on the volume fraction $\phi$.
We also offer evidence that the chaotic motion is responsible for the loss 
of memory in the evolution of the system and 
demonstrate this loss of correlation in phase space. 
The loss of memory at the microscopic level of individual particles 
is also shown in terms of the autocorrelation 
functions for the two transverse velocity components.
Moreover, a negative correlation in the transverse particle velocities
is seen to exist at the lower concentrations, an effect which we explain 
on the basis of the dynamics of two isolated spheres undergoing simple shear.
In addition, we calculate
the probability distribution function of the velocity fluctuations
and observe, with increasing $\phi$, a transition from exponential 
to Gaussian distributions. 

The simulations include a non-hydrodynamic repulsive
interaction between the spheres which qualitatively models the effects of
surface roughness and other irreversible effects, such as residual Brownian
displacements, that become particularly important whenever
pairs of spheres are nearly touching. We investigate the effects of 
such a non-hydrodynamic interparticle force on the scaling 
of the particle tracer diffusion coefficient $D$ for very dilute suspensions,
and show that, when this force is very short-ranged,
$D$ becomes proportional to $\phi^2$ as $\phi \to 0$.
In contrast, when the range of the non-hydrodynamic
interaction is increased, we observe a crossover in the dependence of $D$ on $\phi$, 
from $\phi^2$ to $\phi$ as $\phi \to 0$.  
\end{abstract}

\section{Introduction}
\label{intro}

The phenomenon of shear-induced particle diffusion in suspensions at vanishingly small 
Reynolds number has been studied extensively since the work of \cite{eckstein77}, where 
it was first suggested that a shear flow causes the particles to execute random migrations
across the streamlines of the ambient flow producing an effect akin to dispersion. 
This hypothesis of a self-diffusive motion arising from purely viscous hydrodynamic 
interactions between particles has already received considerably experimental support
\cite[][]{eckstein77,leighton87a,breedveld98,breedveld2001a,breedveld2001b}.
Moreover, as pointed out by \cite{breedveld2001a}, the origin of this diffusive behaviour is 
clearly different from the more familiar Brownian diffusion in colloidal suspensions caused 
by thermal fluctuations, as well as turbulent diffusion driven by inertial effects, in that 
this shear-induced diffusion is due only to the hydrodynamic interactions between the particles 
comprising the suspension. Since, in principle, these interactions constitute a deterministic 
process, the question arises as to whether and how it can also be viewed as a diffusion process. 
To address this question, it is generally assumed that the arrangement of neighboring suspended 
spheres leads to a time-dependent random event \cite[][]{eckstein77} in that, upon {\it collision} 
with their closest neighbors, the spheres will suffer many successive random displacements 
ultimately leading to a random walk \cite[][]{leighton87a,zarraga99}.
Underlying this description is the fundamental assumption that collisions between 
spheres eventually become statistically independent.
Although many theoretical studies have made this strong assumption in order to calculate the 
diffusivity of very dilute sheared suspensions from the displacement produced by a single 
collision and then averaging over all initial configurations of the colliding spheres
\cite[][]{acrivos92,wang96,wang98,cunha96}, the complex dynamics of suspensions undergoing shear, 
leading to the loss of correlations in the particle motions, has not been investigated thus far
in much detail. 

It is the purpose of the first part of this work to pursue the recent suggestion made by 
\cite{marchioro2001} that the chaotic evolution of sheared suspensions is responsible for 
the {\it loss of memory} referred to above, and to demonstrate, via numerical simulations, 
that the evolution of the system in phase space is indeed chaotic. Recall that chaotic 
dynamics is characterized by the sensitivity of the system to initial conditions, as evidenced 
by the exponential growth of the separation distance in phase space of two initially neighboring 
trajectories, and that a standard measure of this sensitivity is the largest Lyapunov exponent 
(LLE), which gives the average rate of this separation distance \cite[][ p. 85]{baker}. 
We shall therefore investigate the LLE as a function of the volume fraction $\phi$ 
of the suspensions. We shall also show that the system loses the memory of its initial 
state and that its evolution is asymptotically diffusive. The loss of memory 
at the level of a single sphere will also be discussed in terms of the autocorrelation 
functions of the two transverse velocity components, and we will show that, 
as the concentration is increased and collisions between spheres 
become more frequent, the time scale over which the particle transverse velocities
remain correlated is shortened. 

Before proceeding, it is instructive to consider the motion of a tracer sphere 
subject to purely hydrodynamic interactions with the other spheres in the
suspension. As is well-known, and as 
a direct consequence of the linearity of the equations 
of motion at zero Reynolds number, in any encounter between 
two perfectly smooth spheres neither sphere experiences a net lateral
displacement, although both may suffer large transient displacements 
from their original streamlines \cite[][ p. 257]{leal}. Therefore,  for the tracer to
experience a net displacement leading to diffusive motion it is
necessary that it interact with at least two other spheres.
Since the rate of simultaneous interactions of a tracer sphere with two
other spheres is proportional to $\gamma \phi^2$ as
$\phi \to 0$, where $\gamma$ is the shear rate, and since the magnitude of each 
displacement is proportional to the particle radius $a$, the self-diffusion
coefficient should be proportional to $\gamma \phi^2 a^2$,
in the limit of very dilute suspensions \cite[][]{leighton87a}.
The experimental results by \cite{leighton87a} observe this scaling for 
volume fractions $0.05 < \phi < 0.40$. 
However, in any real experiment, suspended particles are not perfectly
spherical and, as the separation between two colliding particles can be 
very small (less than $10^{-4}$ of a particle radius \cite[][]{cunha96}), 
even very small asperities might play a significant role during 
encounters between a pair of spheres.
In experiments performed by \cite{rampall97} a roughness of order $10^{-3}$
particle radii was found, and the effect of the asperities 
on the trajectories of nearly touching spheres during a collision was determined. 
In this situation, the interaction of the tracer 
particle with another sphere {\em will} lead to a net displacement of the tracer
from its original streamline, and therefore, since the rate of interactions
with another sphere is proportional to $\gamma \phi$, the diffusion coefficient
is expected to scale as $\gamma \phi a^2$ as $\phi \to 0$.
This linear dependence of the diffusion coefficient on $\phi$ was
observed in experiments performed by \cite{phan99} and by \cite{zarraga99}. 
On the other hand, thus far, the numerical simulations of sheared suspensions
have not been extended to low enough values of $\phi$ where one 
or the other of these two regimes for the diffusivity would be expected 
to apply. 

In this work, we shall therefore investigate the scaling of the
diffusion coefficient by performing simulations down to values of 
$\phi$ as low as $0.03$. The effect of surface 
roughness and other possible non-hydrodynamic forces, such as residual Brownian forces,
will be qualitatively modelled by introducing a short-ranged repulsive force between
the spheres. This is similar to the approach taken by
\cite{cunha96} and by \cite{zarraga99}, where the surface roughness was modelled as
a normal force that prevents the particles from coming closer than a 
certain fixed distance. As discussed above, the contribution to the diffusive
motion of the spheres due to binary collisions clearly depends on the magnitude 
and range of the interparticle force, in that for a weak force, a 
$\gamma \phi^2 a^2$ regime for the diffusivity would be expected. 
However, since at low enough concentrations,
the linear regime should eventually become dominant 
no matter how small the effect of the interparticle force, 
we shall investigate this transition
from a quadratic to a linear dependence of the diffusivity on $\phi$ as
$\phi \to 0$. As a preliminary, in \S\,\ref{interparticle_force} we shall discuss the effects
of such a non-hydrodynamic interparticle force on the microscopic structure of the suspension,
and in particular its dependence on the strength and the range of this force. 

\section{Simulation method: Stokesian dynamics}
\label{simulation}

We consider suspensions of non-Brownian particles undergoing shear using
the method of Stokesian Dynamics which was  specifically developed for dynamically 
simulating the behaviour of many particles suspended in a fluid \cite[]{brady84}.
A detailed description of the method is given 
in a review by \cite{brady88}, hence only a brief discussion is presented here.
The method accounts for both hydrodynamic
and non-hydrodynamic forces between particles. 
Hydrodynamic forces are computed for spherical particles undergoing 
simple shear, characterized by a shear rate $\gamma$, in the limit of
zero Reynolds number. In simulating the behaviour of infinite suspensions, 
periodic boundary conditions in all direction are imposed, using an adapted
version of the Lees-Edwards boundary condition in the direction of the shear 
\cite[][ p. 242]{allen}\cite[]{brady85}. The volume $V$ of the cubic cell
containing a fixed number of spheres $N$ is related to the
volume fraction $\phi$ by $\phi=(4\pi a^3/3) N/V$.  
Interactions between particles
more than a cell apart from each other cannot be neglected,
due to the long-range character of the hydrodynamic forces, and
a lattice sum of the interactions, using the Ewald method, is 
implemented \cite[]{brady88b}. 

A typical simulation here consists of $N=64$ particles 
sheared over a period of time $t \sim 100 \gamma^{-1}$.
The results are averaged over $N_c \sim 100$ different initial 
configurations using the random-phase average method proposed 
by \cite{marchioro2001}, in order to avoid spurious time-periodic 
fluctuations induced by the time-dependent shape of the simulation 
cell. Each initial configuration corresponds to a random
distribution of non-overlapping spheres in the simulation cell. 

In what follows, we shall express all the variables in dimensionless units,
using the radius of the spheres $a$ as the characteristic length and
$\gamma^{-1}$ as the characteristic time.

\subsection{Non-hydrodynamic interparticle force}
\label{interparticle_force}

In a suspension of non-Brownian spherical particles undergoing shear
at zero Reynolds number, the separation between spheres can be very small
(less than $10^{-4}$ of their radius). In this situation, the effects of
surface roughness or small Brownian displacements cannot be neglected and
a short-ranged, repulsive force is usually introduced between the spheres 
to qualitatively model the behaviour of real systems. The introduction of such a force
has the further numerical advantage of preventing overlaps between the spheres. 

\begin{figure}
\centering
\includegraphics[width=10cm,angle=-90]{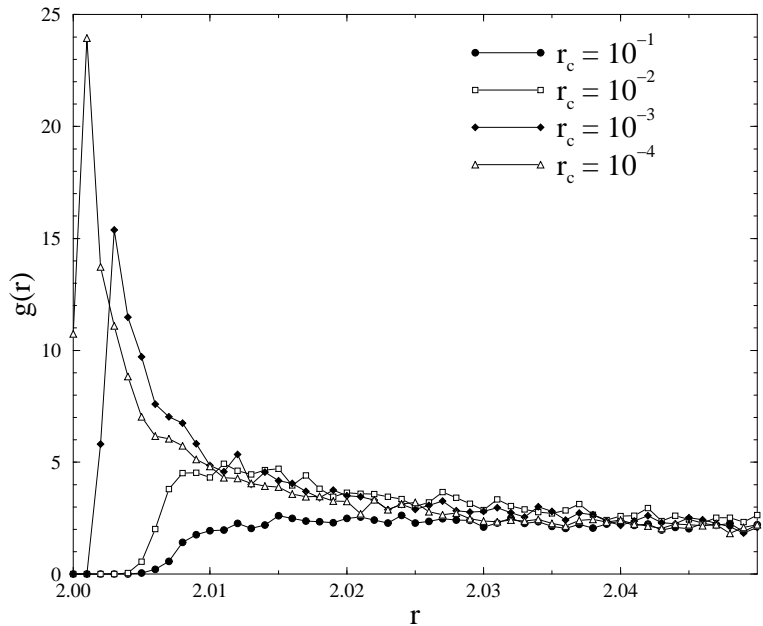}
\caption[gr]{First peak of the pair distribution function $g(r)$ 
for different values of $r_c$, the characteristic range of the 
interparticle force. The position of the peak indicates that,
as the range of the repulsive force increases, so does
the minimum separation between particles. 
The simulations are for $\phi=0.10$, $N=64$, $F_0=1.0$, and $N_c=100$.
$g(r)$ is measured after each initial random distribution of spheres
is sheared for $t \sim 50$ and steady state was reached.}
\label{gr}
\end{figure}

In this work we used the expression for the repulsive interparticle force, 
already well-tested in the context of Stokesian dynamics,
\begin{equation}
\label{inteparticle}
{\bf F}_{\alpha \beta}= \frac{F_0}{r_c}
\frac{{\rm e}^{-\epsilon/r_c}}{1- {\rm e}^{-\epsilon/r_c}} 
{\bf e}_{\alpha \beta} \, ,
\end{equation}
where 6$\pi \mu a^2 \gamma{\bf F}_{\alpha \beta}$,
with $\mu$ being the viscosity of the suspending liquid,
is the force  exerted on sphere $\alpha$ by sphere $\beta$,
$F_0$ is a dimensionless coefficient reflecting the magnitude of this force,
$r_c$ is the characteristic range of the force,
$\epsilon$ is the distance of closest approach between the surfaces
 of the two spheres divided by $a$,
and ${\bf e}_{\alpha \beta}$ is the unit vector connecting their centres
pointing from $\beta$ to $\alpha$.

The magnitude and range of the interparticle force affect the
microstructure of the suspension \cite[][]{brady97}. As mentioned earlier, colliding 
spheres in a shear flow almost touch one another, with the actual 
minimum separation distance strongly dependent on the 
repulsion between them. Following \cite{bossis84} we show this effect
in figure \ref{gr} in terms 
of the pair distribution function $g(r)$, defined as the probability of
finding the centre of a second particle at a distance $r=|{\bf r}|$ given that
there exists a sphere with its origin at ${\bf r}=0$. We see that the
minimum separation, and therefore the first peak in $g(r)$, is
strongly affected by the range of the interparticle force in that,
as $r_c$ increases, the minimum separation between neighboring particles 
also increases.

\begin{figure}
\centering
\includegraphics[height=12cm,angle=-90]{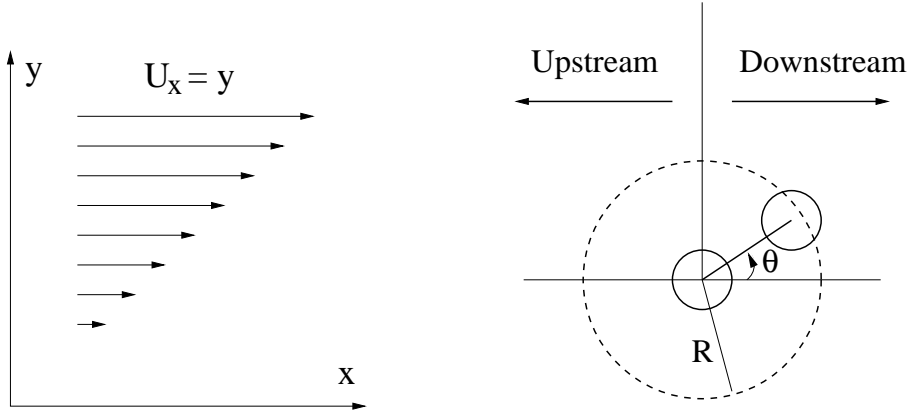}
\caption[gr]{Diagram of a pair of spheres oriented with an angle
$\theta$ measured from the downstream side of the reference sphere.
Down and upstream sides of the reference sphere as well as the distance $R$
defining particle pairs are shown.}
\label{orientation}
\end{figure}

\begin{figure}
\centering
\includegraphics[width=10cm,angle=-90]{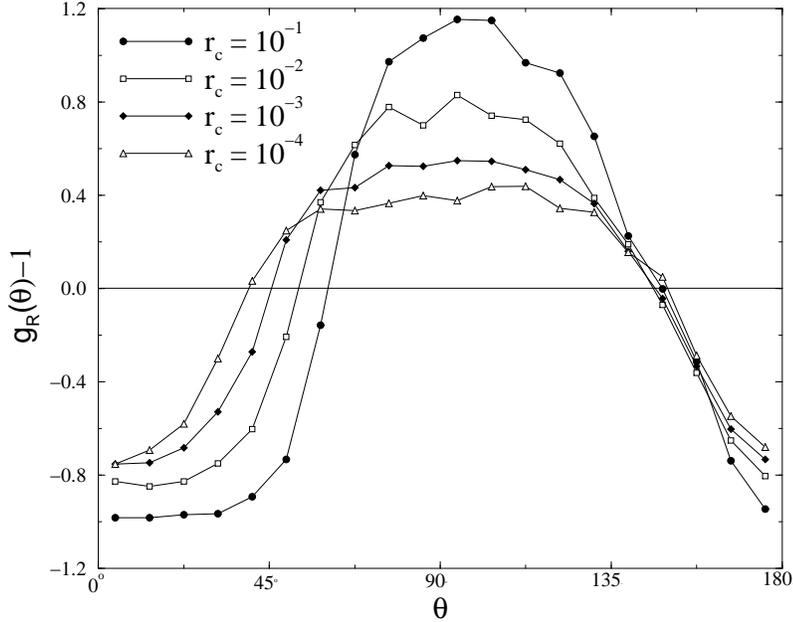}
\caption[gr]{Normalized angular distribution function $g_{R}(\theta)$
for pairs of particles. The distance between the particles is $2<r<2.1$ 
($R =2.1$). Different curves correspond to different values of
the interparticle force range $r_c$. All simulation were performed with
$N=64$, $N_c=100$, $F_0=1.0$ and $\phi=0.10$.}
\label{gr_theta}
\end{figure}

It is also known that the presence of a repulsive force
breaks the angular symmetry of the microstructure and,
in particular, that it destroys the fore-aft symmetry of the particle
trajectories in a simple shear flow \cite[][]{bossis84,dratler96}.
Specifically, again following \cite{bossis84}, 
let us consider the angular orientation of {\it pairs}
formed by spheres closer than a certain distance $R$, and let us
define  the probability distribution $g_{R}(\theta)$
as the probability density of finding a pair having a given 
orientation angle $\theta$ (c. f. figure \ref{orientation}).
It is clear that, as already shown by \cite{bossis84}, 
particle pairs would be expected to spend more time 
oriented on the upstream side, where the repulsive force is balanced by the
shear forces pushing the two particles together, implying that 
the probability of finding a pair oriented upstream, $90^\circ<\theta<180^\circ$, 
would be higher than in the downstream range, $0^\circ<\theta<90^\circ$. 
This is borne out by figure \ref{gr_theta} which shows that the
pair distribution function becomes increasingly more asymmetric,
favoring the upstream orientation of particle pairs, as the
range of the interparticle force is increased.

In addition to the microscopic effects mentioned above, the interparticle force
also affects macroscopic measurable quantities, as we shall
discuss in the following sections.

\section{Chaotic motion and time-reversibility}
\label{chaos}

As already mentioned in the introduction, the question as to whether 
and how diffusive-like transport arises in a 
suspension of non-Brownian particles undergoing shear has been investigated 
since the original work by \cite{eckstein77}.
Experimental evidence strongly suggests that even with vanishingly small 
inertia effects (zero Reynolds number) and negligible Brownian and 
non-hydrodynamic forces, sheared suspensions exhibit diffusive behaviour
\cite[][]{leighton87a,breedveld98,breedveld2001a,breedveld2001b}.

In an ideal case, where only hydrodynamic forces are present and 
the Reynolds number is exactly zero, the motion of the particles 
is deterministic and reversible due to the linearity of the governing flow equations,
thereby implying that, upon reversing the direction of flow, the particles 
should retrace their trajectories. However, in any physical experiment, 
there exists inherent slight irreversible effects at the microscopic level,
such as  residual Brownian motion and surface-roughness effects, which,
as is usually the case in dynamical systems, can have a strong impact on 
macroscopic measurable quantities.
As an example, let us consider again the loss of 
fore-aft symmetry in sheared suspensions.

It can be shown, based upon the reversibility of Stokes flow,
that the pair distribution function of sheared suspension of 
perfect spheres should have fore-aft symmetry.
On the other hand, since the original work by \cite{gadala80},
there exists strong experimental evidence that, even at 
vanishingly small Reynolds numbers and Brownian force effects, sheared
suspensions develop an anisotropic structure resulting in the loss
of fore-aft symmetry \cite[][]{parisi87,rampall97}. This broken symmetry
has been attributed to surface-roughness effects. As 
mentioned in \S\,\ref{interparticle_force} the
minimum separation between two colliding spheres in a shear flow can be 
less than $10^{-4}$ of their radius, so that even a small surface roughness
can have an important influence in this case \cite[][]{cunha96}. 
Thus, small irreversible effects related to surface roughness, which is
present at microscopic scales, have a measurable impact on the
macroscopic structure of the suspension and correspondingly on related
macroscopic quantities such as the normal stress difference in
sheared suspensions. 

Thus far we have discussed the microscopic origin 
of irreversibility and its manifestation in macroscopic quantities which 
provides a microscopic basis for the existence of an intrinsically irreversible
macroscopic description as given by the diffusion equation.
But, the basic assumption underlying the derivation of such an equation and,
in particular, the validity of a statistical description of the system
based on the randomness of the microscopic motion of the particles, is
in need of further discussion. 
Recall that, in the calculation of the diffusivity in 
very dilute sheared suspensions, it is generally assumed that 
successive collisions between spheres are statistically independent, 
sometimes referred to as {\it molecular chaos} 
\cite[][]{acrivos92,wang96,wang98,cunha96}. 
Furthermore, it has been suggested that a close connection 
exists between {\it molecular chaos} and dynamical chaos, 
according to which stochastic-like behaviour is possible 
even for deterministic mechanical systems 
(e. g. \cite{gaspard}, p. 225; \cite{dorfman98}.) 
In fact, even low-dimensional deterministic dynamical systems are 
known to give rise to diffusive transport (usually named 
{\it deterministic diffusion} \cite[][ p. 293]{gaspard}).
Chaotic systems have the property that any small perturbation
in the state of the system will grow exponentially
in time, to the point where the evolution can no longer be
accurately predicted. But, although the hypothesis of chaotic motion 
as the basic mechanism responsible for this 
{\it loss of memory} and for that matter, as the cause of the phenomenon of
shear induced diffusion, has been suggested in the context
of sheared suspensions by \cite{marchioro2001} and has been supported
by their numerical simulations showing irreversible 
behaviour upon reversal in the direction of the shear, to-date the  
presence of chaos in a sheared suspension has 
not been examined. We shall therefore numerically investigate 
the presence of chaotic motion in sheared suspensions by
evaluating  the largest Lyapunov exponent which is a standard
measure of chaoticity \cite[][ p. 24]{schuster}. 

It is worth mentioning here that we do not wish to imply that
diffusive motion cannot possibly operate in the purely hydrodynamic case,
and that diffusion owes its existence to the unavoidable presence
of small microscopic irreversible forces.
In fact, the presence of chaotic motion, which we shall presently demonstrate,
strongly suggests that, even in the absence of such effects, diffusive 
behaviour should still arise due to the loss of correlation in
the particle motions.

\subsection{Largest Lyapunov exponent}
\label{lyapunov}

In a sheared suspension at zero Reynolds number,
the state of the system is fully determined
by the coordinates of all the particles and, 
therefore, a sheared suspension consisting of $N$ particles 
can be considered as a dynamical system in a $3N$-dimensional
phase space. A point of this phase space ${\bf \Gamma}$ is given
by $3N$ particle coordinates and the 
Lyapunov exponents measure the average rate of separation of two
initially neighboring trajectories in phase space.
Thus, given two states of the system 
separated a small distance $d(0)$ in phase space, 
the largest Lyapunov exponent (LLE) for that state 
is formally defined as 
\begin{equation}
\label{lyapunov_formal}
\lambda = \lim_{t\to \infty} \lim_{d(0) \to 0} 
\left[ \frac{1}{t} \frac{d(t)}{d(0)}\right],
\end{equation}
with $d(t)$ being the separation distance in phase space at time $t$,
which controls the exponential divergence of initially close trajectories.

\begin{figure}
\centering
\includegraphics[width=10cm,angle=-90]{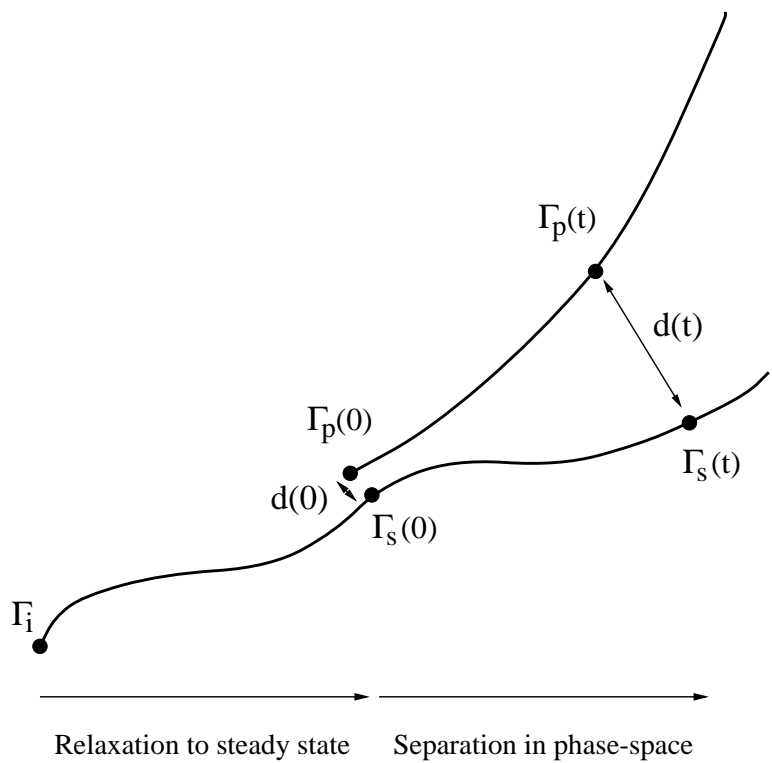}
\caption[phase]{Schematic representation of the generation of
initially close trajectories in a stationary state. Starting from 
a random configuration of hard-spheres ${\bf \Gamma}_i$ we follow the evolution 
of the system towards steady state. Once in steady state, we generate
a slightly perturbed state ${\bf \Gamma}_p$ by adding a small random
displacement to each particle. The initial separation in
phase space is $d(0)$. Then we follow the evolution of the
two systems by computing the distance at each time $d(t)$.}
\label{phase-space}
\end{figure}

In order to obtain a numerical estimate of the LLE for the steady state of 
a sheared suspension, we compute the ensemble averaged LLE
used in ergodic systems \cite[][ p. 144, 247]{gaspard} by generating a number $N_c$ of 
initial configurations ${\bf \Gamma}_{s}(0)$ in a steady state flow, 
starting from a random configuration ${\bf \Gamma}_i$ of spheres and evolving 
the system for a sufficiently long time (typically for strains $t \sim 50$). 
For each of these configurations we then introduce a slightly perturbed state 
${\bf \Gamma}_{p}(0)$, by adding a random displacement $\bar{d}$
to ${\bf \Gamma}_{s}(0)$ \footnote{In order to generate a 
random perturbation in phase space $\bar{d}$, with constant magnitude 
$|\bar{d}|=d(0)$, we initially construct a  3N-dimensional vector 
$\bar{\xi}$, where each component $\xi_i$ is a random variable
with uniform distribution in the range $-1 < \xi_i <1$. Then, we renormalize
this vector to obtain the desired initial magnitude of the perturbation,
$\bar{d}=\bar{\xi} \, (d(0)/|\bar{\xi}|)$. Thus, a random displacement $\bar{d}$ 
in phase space corresponds, on average, to a random perturbation, 
of order $d(0)/\sqrt{N}$, to the position of each particle.}
(see figure \ref{phase-space}). 

\begin{figure}
\centering
\includegraphics[width=10cm,angle=-90]{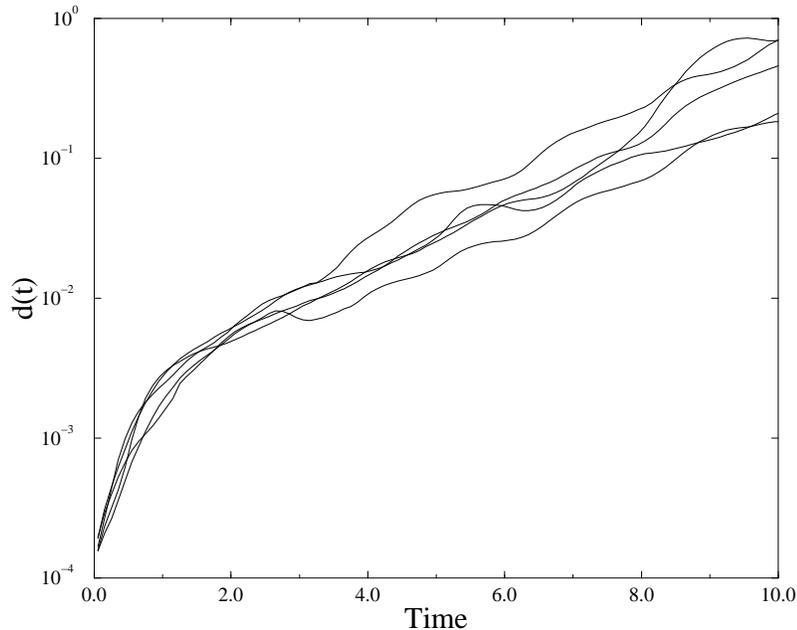}
\caption[phase]{Separation of initially close trajectories in phase space
$d(t$) for different initial states of the unperturbed system. 
The results correspond to simulations with $N=64$, $F_0=1.0$, $r_c=10^{-4}$, 
and $\phi=0.35$. }
\label{single}
\end{figure}

Following the evolution of these two independent systems 
we can compute a distance in phase space given by,
\begin{equation}
\label{d}
d(t) = || {\bf \Gamma}_s(t) - {\bf \Gamma}_p(t) || = 
\left( \sum_{i=1}^{N} 
\left[ {\bf x}_s^i(t) - {\bf x}_p^i(t)\right]^2 
\right)^{\frac{1}{2}}
\end{equation}
where the indices $s$ and $p$ refer to the unperturbed and
perturbed initial states of the two systems\footnote{Let us mention that, 
even though the affine shearing motion of the spheres is not removed and its 
contribution dominates the initial growth of the separation distance 
in phase space, the exponential growth should eventually become dominant
at long times.} (see figure \ref{phase-space}). 
To be sure, for a single realization, $d(t)$ will depend on both initial states
 ${\bf \Gamma}_s(0)$ and ${\bf \Gamma}_p(0)$, 
but, after some transient behaviour, we should certainly expect
an exponential separation with large fluctuations
due to the details of the dynamics of the system. This is illustrated in figure 
\ref{single}. In order to compute the LLE accurately, we therefore first average over 
$N_c$ different initial conditions in phase space, then
take the logarithm of the mean exponential separation,
\begin{equation}
\label{dx-average}
\Delta_\Gamma(t) = \ln{\left(\langle d(t) \rangle_{{\bf \Gamma}}\right)}
= \ln \left( \frac{1}{N_c} \sum_{k=1}^{N_c} \{ d(t) \}_k \right)
\end{equation}
and identify $\lambda_\Gamma$ (the largest Lyapunov exponent) as the slope
of $\Delta_\Gamma(t)$ in the region of its linear growth
\footnote{The method used in this work to compute the largest Lyapunov 
exponent differs from that described by \cite{benettin76} and used frequently. 
Instead of averaging over different
realizations (ensemble average), \cite{benettin76} reset the 
distance between trajectories to the initial value after 
fixed intervals of time $\tau$ and compute the Lyapunov exponent from the average
distance reached before the rescaling procedure (time average). 
However, both methods should give the same value of $\lambda$ if
the system is ergodic, because in that case, the time-average of any dynamical 
quantity is equal to its ensemble average over a large number of
realizations $N_c$ \cite{ruelle85}.
The main advantage of our method is that it
allows us to compute the evolution of all the initial configurations
simultaneously, which reduces the computational time enormously,
as compared to a single very long simulation, when the computations
are performed using a cluster of processors.}.

\begin{figure}
\centering
\includegraphics[width=8.7cm,angle=-90]{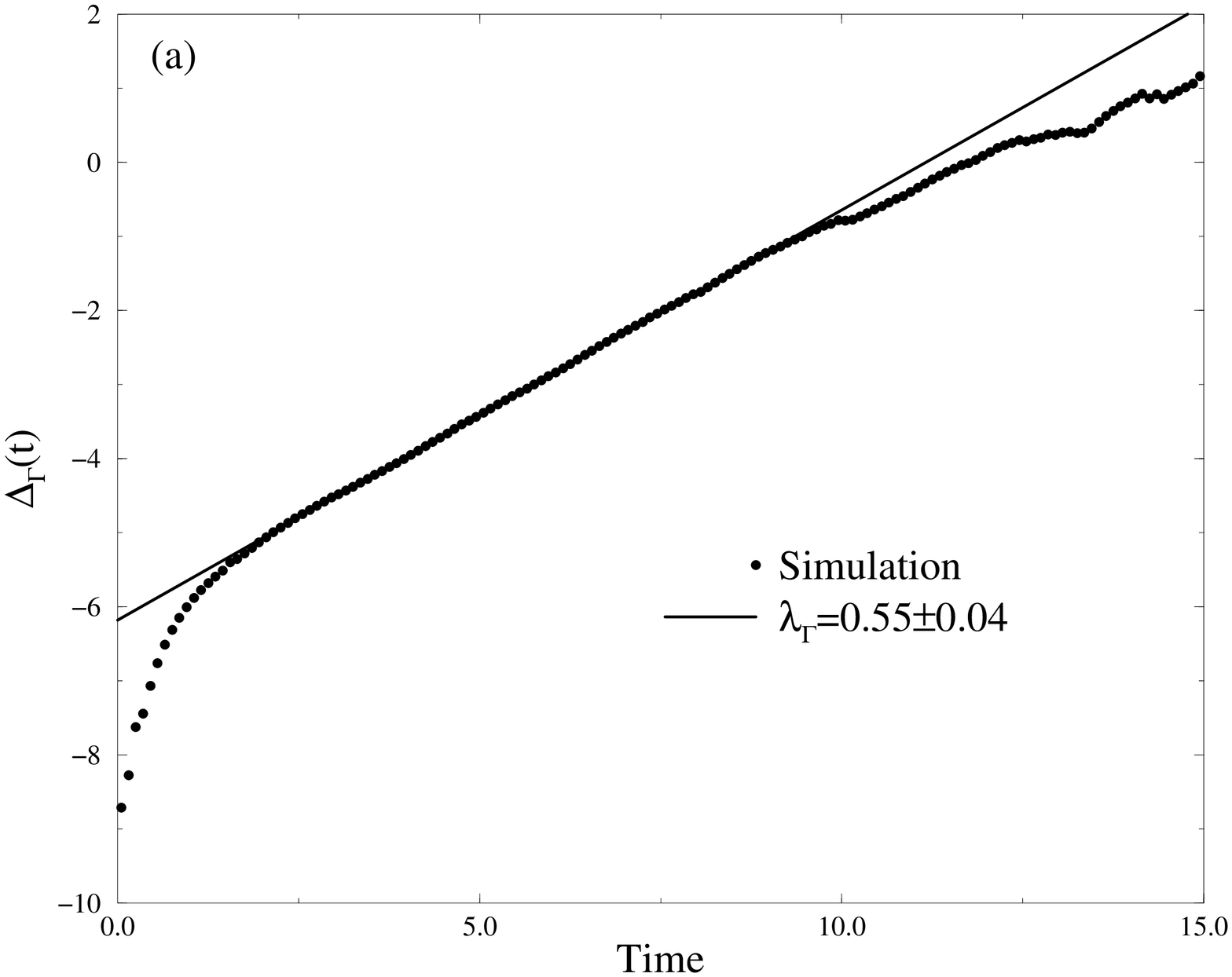}
\includegraphics[width=8.7cm,angle=-90]{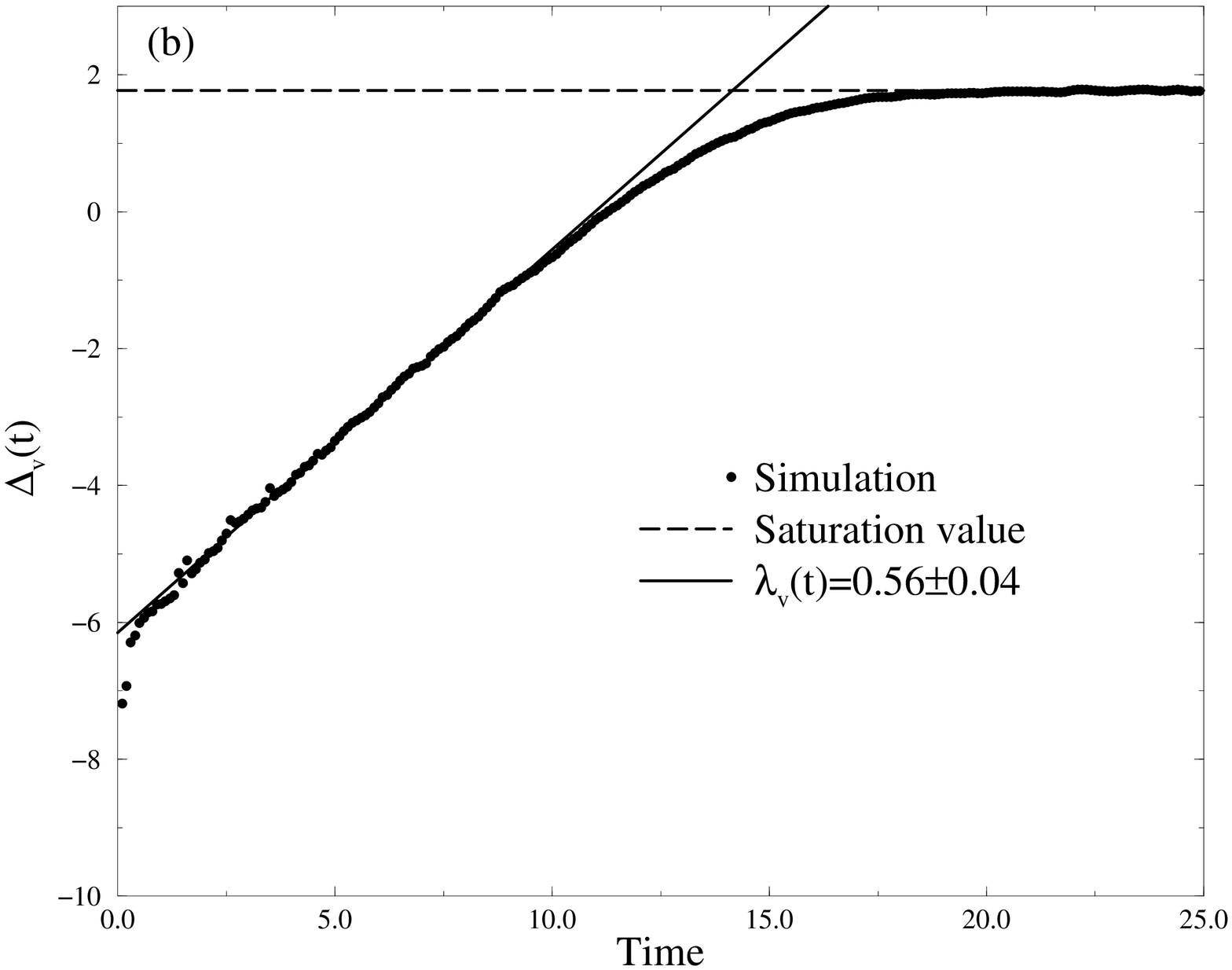}
\caption[distance]{Separation of initially close trajectories in
phase space. (a) Coordinate phase space, (b) Transverse velocity space.
The results corresponds to a sheared suspension of $N=64$
particles, volume fraction $\phi=0.35$, interparticle force $F_0=1.0$,
characteristic range $r_c=10^{-4}$, and $N_c=100$. The solid line corresponds
to a linear fit with (a) $\lambda = 0.55 \pm 0.04$, 
(b) $\lambda = 0.56 \pm 0.04$. The dashed line in (b) is the 
asymptotic value of $\Delta_v$ as $t \to \infty$ given by 
(\ref{saturation}).}
\label{distanceXV}
\end{figure}

Analogously, we can define a distance in the space of
the transverse velocity components of all the particles,
\begin{equation}
\label{dv-average}
\Delta_v(t)
= \ln \left[ \frac{1}{N_c} \sum_{k=1}^{N_c}
\left\{
\left( \sum_{i=1}^{N}
\left[ {\bf v}_s^i(t) - {\bf v}_p^i(t)\right]^2
\right)^{\frac{1}{2}}
\right\}_k
\right]
\end{equation}
However, as $d(t)$ grows exponentially, so does a generic projection in phase space
(or a distance measured with any particular metric) 
Moreover, since the separation is dominated 
by the exponential growth with the LLE, and the velocity is the
derivative of the position of the particles, one would expect
the Lyapunov exponent $\lambda_v$, as obtained from the slope
of $\Delta_v(t)$, to be the same as that measured from
$\Delta_\Gamma(t)$, i.e. $\lambda=\lambda_\Gamma=\lambda_v$. 

In figure \ref{distanceXV} we show the time evolution of $\Delta_\Gamma(t)$ 
and $\Delta_v(t)$ for a sheared suspension with 
$N=64$ particles, volume fraction $\phi=0.35$, interparticle force $F_0=1.0$ and
characteristic range $r_c=10^{-4}$.
The distance $d(t)$ was averaged over $N_c$=100 different initial 
conditions. The initial distance in phase space ${\bf \Gamma}$ 
is $d(0)=10^{-4}$, which corresponds to a random displacement
$\sim 10^{-5}$ added to the position of each particle. 
Simulations using a smaller initial displacement
give a similar behaviour with a variation in the measured Lyapunov
exponent less than $5\%$. 
We also performed simulations with $r_c=10^{-3}$ as well as with $F_0=0.1$,
but, in all cases, the variations between the LLE's
were within $10\%$. However, larger changes should be expected in the
value of the LLE for larger variations in the range or the 
strength of the interparticle force. Finally, the Lyapunov exponent calculated from 
$\Delta_\Gamma(t)$ and from $\Delta_v(t)$ is, as expected, 
the same, within the error in its determination.

In figure \ref{distanceXV} three different regimes can be directly observed,
a short initial transient behaviour, a large linear growth corresponding to
the exponential divergence in phase space and a deviation from the
linear growth at long times, corresponding to an asymptotic 
saturation in velocity space. 

There are two related effects that
might cause the short-time transient behaviour.
On one hand, many-body nonlinear effects only come into play after at least 
one collision has taken place, and we can therefore relate the 
transient time to a characteristic time between collisions. 
During this initial time the distance grows mainly due to the shear flow
imposed to the particles. However, the definition of such a 
characteristic time is not clear, given the fact that the hydrodynamic 
forces are long ranged. 
The experiments of \cite{breedveld2001a} found a similar transition 
when measuring the diffusivity of the particles.
They suggested that this transition might correspond to the 
deformation, due to the shear flow, 
of the initial cloud of particles surrounding a test sphere.
In our simulations we found that this transient time decreases 
with increasing concentration, which is consistent
with the suggestion referred to above.

\begin{figure}
\centering
\includegraphics[width=10cm,angle=-90]{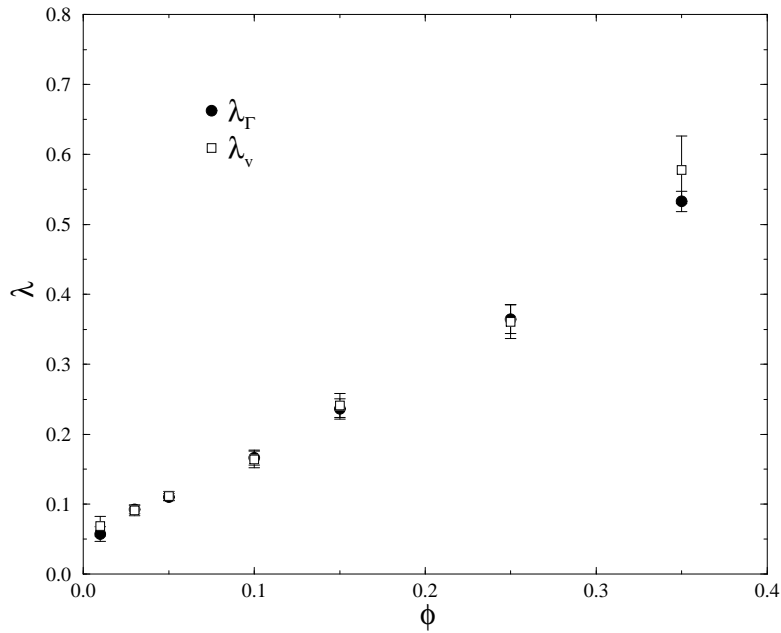}
\caption[lphi]{Dependence of the Lyapunov exponent on the volume fraction.
The results correspond to numerical simulations for $N=64$, $N_c=100$, 
$F_0=1.0$ and $r_c=10^{-4}$. $\lambda_\Gamma$ is computed from the separation 
growth in coordinates phase space, and $\lambda_v$ using the 
transverse velocity space.}
\label{lphi}
\end{figure}

In the simulations it is observed that the distance in the transverse velocity space
saturates. This saturation takes place because the definition of the
distance in the transverse velocity space does not include the affine motion of the particles,
and therefore, two randomly chosen systems in steady state will
have a finite mean distance. In fact, the saturation distance can be 
estimated by assuming that the perturbed system loses memory of its initial 
unperturbed state. In this case, if we further assume that 
$\Delta_v^{\infty}$ becomes
\begin{equation}
\label{saturation}
\Delta_v^{\infty} = \ln \left[
\left\langle
\left[ N
\left\langle
\left( {\bf v}_s^i - {\bf v}_p^i \right)^2
\right\rangle_{N}
\right]^{\frac{1}{2}}
\right\rangle_{{\bf \Gamma}}
\right]
\approx
\ln 
\left[ 
2 N 
\left( 
\sigma^2_{vy} + \sigma^2_{vz}
\right)
\right]^{\frac{1}{2}}
\end{equation}
where $\sigma^2_{vi}$ is the variance in 
the {\it i-th} component of the velocity, the saturation constant can be computed 
in the steady state, independently of the evolution. 
For the simulation shown in figure \ref{distanceXV}, the variances in 
the velocity components are $\sigma^2_{vy}=0.20$ and $\sigma^2_{vz}=0.07$
respectively, giving a saturation value $\Delta_v^{\infty}=1.77$ 
which is in excellent agreement with the observed saturation distance, 
as shown by the dashed line in figure 
\ref{distanceXV}.
On the other hand, as the velocity-space distance saturates, 
the system loses its memory of the initial state so that
the separation in coordinate-phase deviates from the exponential growth,
as can be observed in figure \ref{distanceXV}. In fact, the loss of memory 
of the initial state of the system leads to an 
asymptotic diffusive behaviour in phase space.

\subsection{Dependence of the Lyapunov exponent on the concentration}
\label{l_vs_phi}

We investigated the dependence of the largest Lyapunov exponent
on the volume fraction of the suspension. As the concentration increases,
the frequency of collisions also increases and we may
expect an enlarged sensitivity to the initial conditions.
In figure \ref{lphi} we show the numerical values of the LLE
as a function of $\phi$. It can be seen that the stochasticity of
the system increases with concentration, as measured by $\lambda$.
An almost linear dependence can be also observed, with the surprisingly
fact that the LLE appears to remain finite as $\phi \to 0$.
We remark parenthetically that, in kinetic theory the Lyapunov exponent can be shown 
to be a linear function of the frequency of collisions 
(\cite{gaspard}, p. 8; \cite{dorfman98}), and that
in a sheared suspension, the frequency of collisions is 
roughly $\nu \sim \phi$. However, due to the long-range 
interactions between the particles, it is not clear 
that the same relation between the LLE and $\nu$ will continue to apply in the case of
sheared suspensions; this then is the main difference between the present situation 
and kinetic theory, which is based on the existence of 
short-range interactions between the particles. In fact, the presence of
these long-range hydrodynamic forces might be responsible for
the curious behavior, seen in figure \ref{lphi}, of $\lambda$ at
small volume fractions. Unfortunately, 
the observed initial transient time increases as the concentration 
becomes smaller, thereby limiting the range of concentrations which we were able 
to investigate to $\phi > 0.01$. Thus, the functional dependence of the LLE 
on $\phi$ in the dilute limit remains an open problem.

\begin{figure}
\centering
\includegraphics[width=9.5cm,angle=-90]{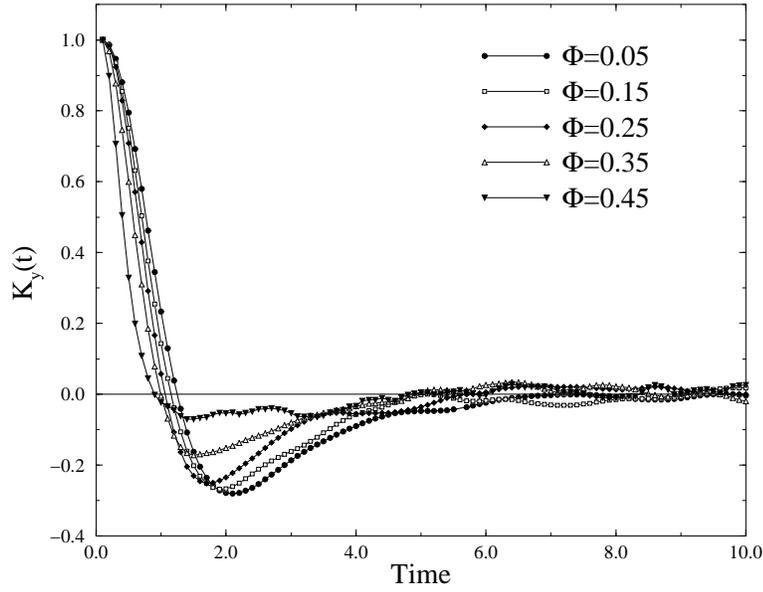}
\includegraphics[width=9.5cm,angle=-90]{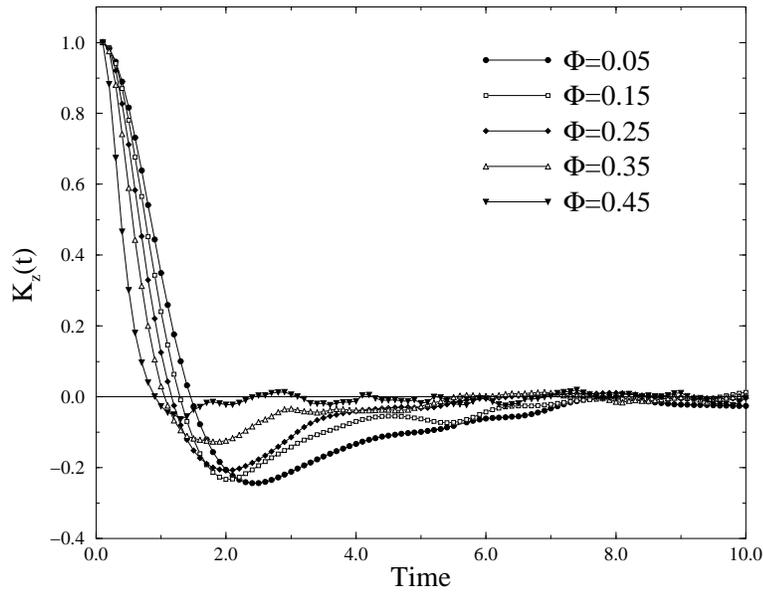}
\caption[correla]{Transverse velocity autocorrelation functions in $y$ and $z$. 
The simulations are for $N=64$, $N_c=100$, $F_0=1.0$ and $r_c=10^{-4}$.}
\label{correla}
\end{figure}

\section{Transverse velocity autocorrelation function}
\label{autocorrela}

In the previous section we discussed the global loss of memory
of a system when its initial state is slightly perturbed.
At the level of individual particles this asymptotic stochastic 
motion is also present, and could be studied by examining
whether the velocity of a particle decorrelates from its
initial value. The average correlations between the velocity of a particle at 
two different times are measured via the velocity autocorrelation function.
Here, we consider $K_{\alpha}(t)$, where $\alpha$ refers to 
the transverse velocity component being 
measured
\begin{eqnarray}
\label{K}
\nonumber
K_{\alpha}(t) &=& 
\langle v_{\alpha}(0) v_{\alpha}(t) \rangle / \sigma_{v_\alpha}^2 
\\  
&=&
\left[ \frac{1}{N_c} \sum_{k=1}^{N_c} \left\{\frac{1}{N}   \sum_{i=1}^{N} 
v^i_{\alpha}(0) v^i_{\alpha}(t)\right\}_k 
\right]
\left/
\left[
\frac{1}{N_c} \sum_{k=1}^{N_c} \left\{\frac{1}{N}   \sum_{i=1}^{N} 
v^i_{\alpha}(0) v^i_{\alpha}(0)\right\}_k  
\right]
\right.,
\end{eqnarray}
and its time integral $\tau_{v_\alpha}$
\begin{equation}
\label{tau} 
\tau_{v_\alpha} =
\frac{1}{\sigma_{v_\alpha}^2} 
\int_0^{\infty} dt \langle v_{\alpha}(0) v_{\alpha}(t) \rangle
= \int_0^{\infty} dt K_{\alpha}(t)
\end{equation}

The importance of these functions
in computer simulations is well known \cite[][ p. 58]{allen} and,
in particular, the velocity autocorrelation function 
is related to the diffusion coefficient through,
\begin{equation}
\label{DvsVV}
D_{\alpha \alpha} = \int_{0}^{\infty} 
\langle v_{\alpha}(0) v_{\alpha}(t) \rangle dt =  
\sigma_{v_{\alpha}}^2 \tau_{v_{\alpha}}.
\end{equation}
This expression has been used in the context 
of suspensions by \cite{nicolai95} to determine the diffusivity 
of sedimenting non-Brownian spheres, and by \cite{marchioro2001} 
to measure the shear-induced self-diffusivities. 

In figure \ref{correla} we show the computed velocity autocorrelation functions
for both transverse velocity components, $K_y(t)$ and 
$K_z(t)$, for volume fractions ranging between $0.05 < \phi < 0.45$, 
where $y$ and $z$ are, respectively, along and normal to the plane of shear.
In all cases, the velocities becomes uncorrelated at long times, 
around $t=10$ for the lowest volume fraction, and from then on
oscillations around zero could be attributed to statistical noise.
Thus, for $t$ beyond about $10$, the displacement of an individual particle is
independent of its displacement at previous times and
hence should be describable 
in terms of a random walk. 

It can also be observed that, except at high concentrations 
($\phi \approx 0.45$), the autocorrelation function becomes
negative over a range of several time units. Let us mention, that a similar 
behaviour for the velocity autocorrelation function was found in molecular 
dynamics simulations of simple liquids \cite[][]{alder58,rahman64}, 
and was explained in terms of the backscattering by neighboring particles at high
densities \cite[][]{alder67,alder70}.
In fact, the negative correlation in velocity suggests that, 
on average, individual particles reverse their velocity,
typically after a time interval of order 1 according to our
simulations. This result can be understood 
in terms of the motion of colliding particles in linear shear flow. 
Specifically, as was already mentioned in the introduction, when two isolated 
spheres collide, the net displacement in both transverse 
directions is zero, as a consequence of the reversibility of 
the creeping flow equations and the symmetry of the problem.
However, the instantaneous deviation in the velocity of the spheres
during the encounter is not zero and clearly anti-correlated, 
since the net displacement should integrate to zero. 
Thus, encounters between two isolated spheres undergoing 
linear shear flow give rise to a negative correlation in
the velocity fluctuations. On the other hand, 
encounters involving three or more particles will, in general,
not be symmetric and the particles will experience a net displacement
from their original streamlines \cite[][]{wang96}. Therefore, 
in general, the interaction between more than two particles 
does not necessarily contribute to a negative autocorrelation in 
the transverse velocity. 

The previous discussion not only explains the observed negative correlation
but also the fact that it becomes more pronounced with decreasing
volume fraction and that, as $\phi \to 0$, it seems to converge
to an asymptotic function dominated by two-particle encounters. 

\begin{figure}
\centering
\includegraphics[width=10cm,angle=-90]{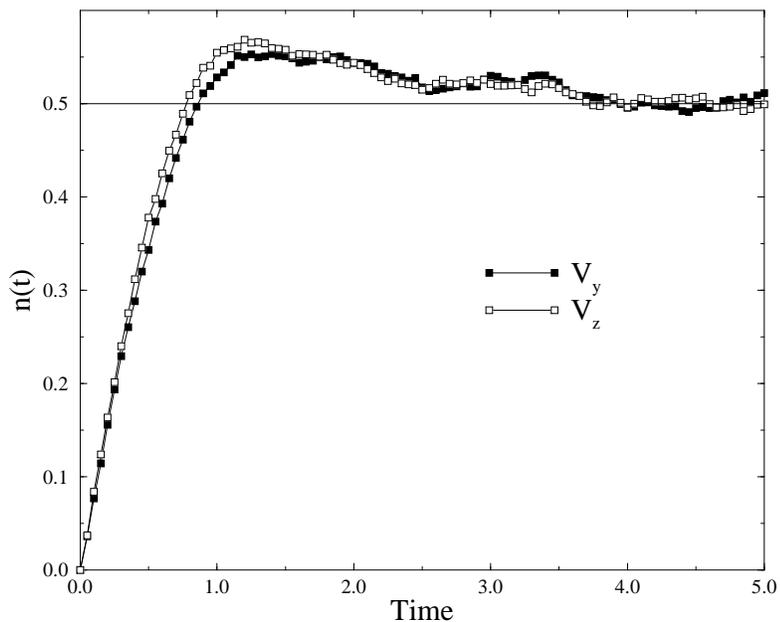}
\caption[inversion]{Average number of particles with their transverse 
velocities reversed from their initial direction. Different curves corresponds to different 
components of the velocity as indicated. 
The simulations are for $N=64$, $N_c=100$, $F_0=1.0$, $r_c=10^{-4}$
and $\phi=0.35$.}
\label{inversion}
\end{figure}

At high concentrations we see that both velocity correlation 
functions decay rapidly to zero and show very little structure.  This
effect might be attributed to `screening': i. e. each of the particles 
alters the ambient velocity field, and when these perturbations 
have a high spatial density and a rapid time dependence, their effect 
is to produce a strongly fluctuating background flow which tends to
decorrelate the velocity of any particle from its earlier value.  An analogous
effect is present in porous media flows, which could be thought of as an
extreme case of a suspension so concentrated as to be immobile.  Here, a static
force perturbation in a fluid-saturated porous medium decays exponentially
with distance (as seen from the Brinkman equation
\cite[][]{brinkman47,durlofsky87b}, for example), in
contrast to the power-law decay in more dilute mobile suspensions.

In figure \ref{inversion} we present the time evolution
of the average fraction of particles having their initial velocities
reversed, 
\begin{equation}
n(t) = \left[1 - 
\left\langle \frac{v_\alpha(0) v_\alpha(t)}{| v_\alpha(0) v_\alpha(t) |} 
\right\rangle \right] \Big/ 2
\label{n}
\end{equation} 
It can be seen that, on average, more than half the particles have
both transverse velocity components reversed at intermediate 
times ($t \sim 1$) for the indicated values of $F_0$, $r_c$ and $\phi$. 
These results are consistent with our previous discussion regarding the 
origin of the negative autocorrelation. 

\begin{figure}
\centering
\includegraphics[width=10cm,angle=-90]{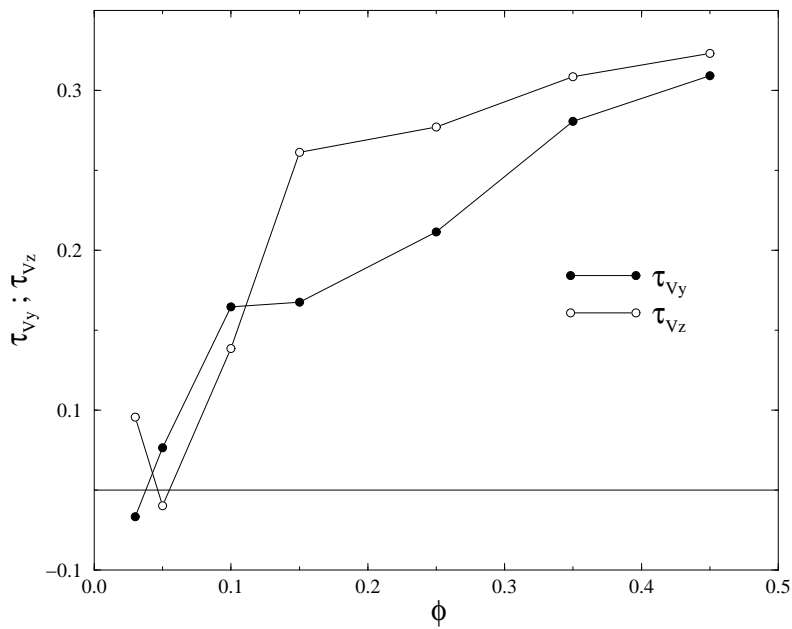}
\caption[correlaT]{$\tau_{v_y}$ and $\tau_{v_z}$ given by equation (\ref{tau}), as a
function of the volume fraction.
The simulations are for $N=64$, $N_c=100$, $F_0=1.0$, and $r_c=10^{-4}$}
\label{correlaT}
\end{figure}

Finally, let us discuss some of the important consequences
of the negative correlations in the transverse velocity 
and their origin. In the first place, it is clear that the 
integral time scale given by equation (\ref{tau})
is different from the autocorrelation time\footnote{A correlation time
defined through equation (\ref{tau}) implicitly assumes 
an exponentially decaying autocorrelation function, as is the case for 
any Gauss-Markov process \cite[][ p. 81]{vankampen}.}. 
This is illustrated in figure \ref{correlaT}, where we depict the values 
for $\tau_{v_{y}}$ and $\tau_{v_z}$, obtained using (\ref{tau}), as 
a function of the volume fraction. 
It is clear that the integral time scale measured from (\ref{tau})
fails to capture the actual time scale underlying
the loss of correlation in the particle velocities in that
it increases with concentration, contrary 
to the fact that the actual time scale for the  correlations becomes shorter 
as the concentration and the rate of collisions between 
the particles increases. Also, given the limited accuracy in the computation of 
the integral (\ref{tau}), the values of $\tau_{v_{y}}$ and $\tau_{v_z}$
shown in figure \ref{correlaT} are subject to large relative errors
when $\phi<0.10$.

The same difficulty is encountered when the diffusion 
coefficient is evaluated by integrating the velocity autocorrelation function. 
As previously discussed, the motion generated by binary collisions
is not diffusive and thus, in view of equation \ref{DvsVV},
the integral of the autocorrelation function should vanish
in the limit $\phi \to 0$. This is born out by the results
shown in figure \ref{correlaT} where it can be seen
that the integral of the autocorrelation function approaches zero 
with decreasing volume fraction. 
Thus, the leading contribution to the diffusivity comes from a
small  portion of the autocorrelation function the role of which becomes
increasingly more crucial as the volume fraction decreases.

\section{Velocity fluctuations}
\label{fluctuations}

\begin{figure}
\centering
\includegraphics[width=9.5cm,angle=-90]{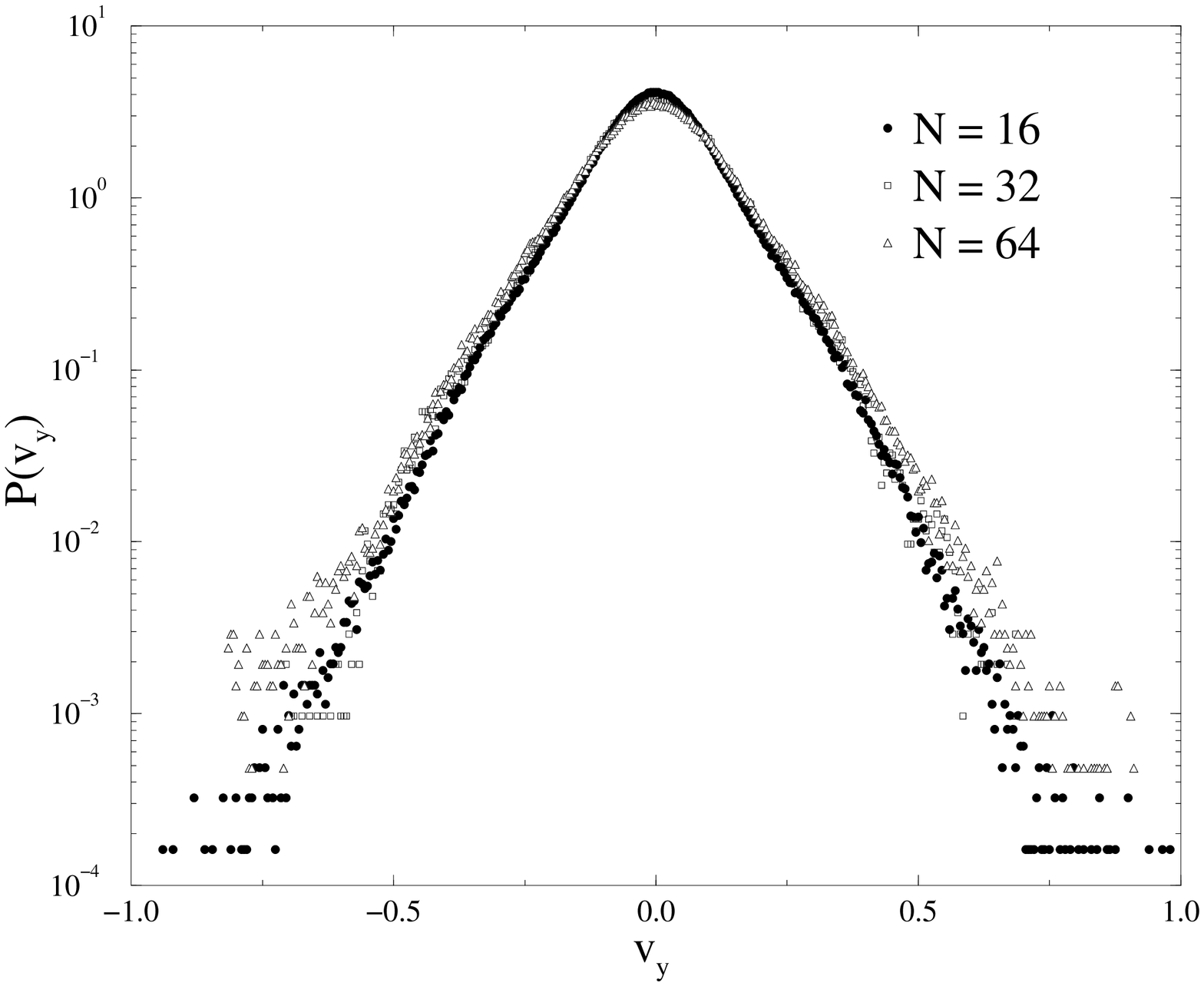}
\includegraphics[width=9.5cm,angle=-90]{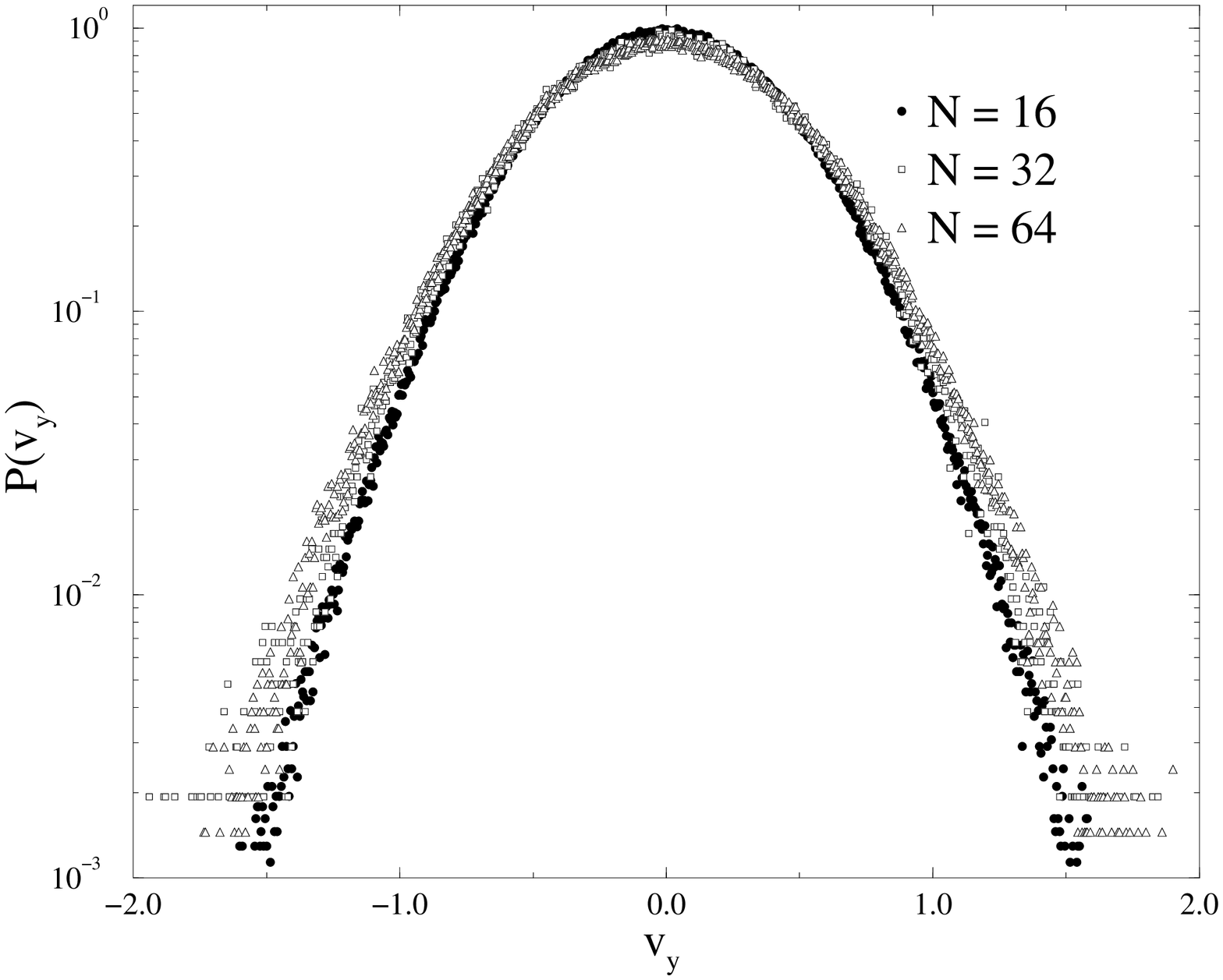}
\caption[pdfN]{Probability density function of the velocity fluctuations
in the direction of the shear $P(v_y)$ for two different volume
fractions $\phi = 0.05$ and $\phi =0.35$. Different curves correspond
to different number of particles in the numerical simulations.
The simulations are for $N=64$, $N_c=100$, $F_0=1.0$ and $r_c=10^{-4}$.}
\label{pdfN}
\end{figure}

In the previous sections we discussed the loss of correlation
in the particle velocities and how the integral time scale
is related to the diffusivity. On the other hand, as is clear 
from equation (\ref{DvsVV}), the magnitude of the velocity fluctuations 
also plays an important role in determining the particle diffusivity. 
Understanding velocity fluctuations in the presence of long-range hydrodynamic 
interactions is a long standing problem,
and has been recurrently studied in the context of
sedimenting suspensions \cite[]{brenner01}. 
In this section we shall present numerical results concerning 
not only the magnitude of the velocity fluctuations given 
by $\sigma_v$ but also the whole probability 
distribution function (pdf) of the velocity fluctuations, 
for different volume fractions.

\begin{figure}
\centering
\includegraphics[width=9.4cm,angle=-90]{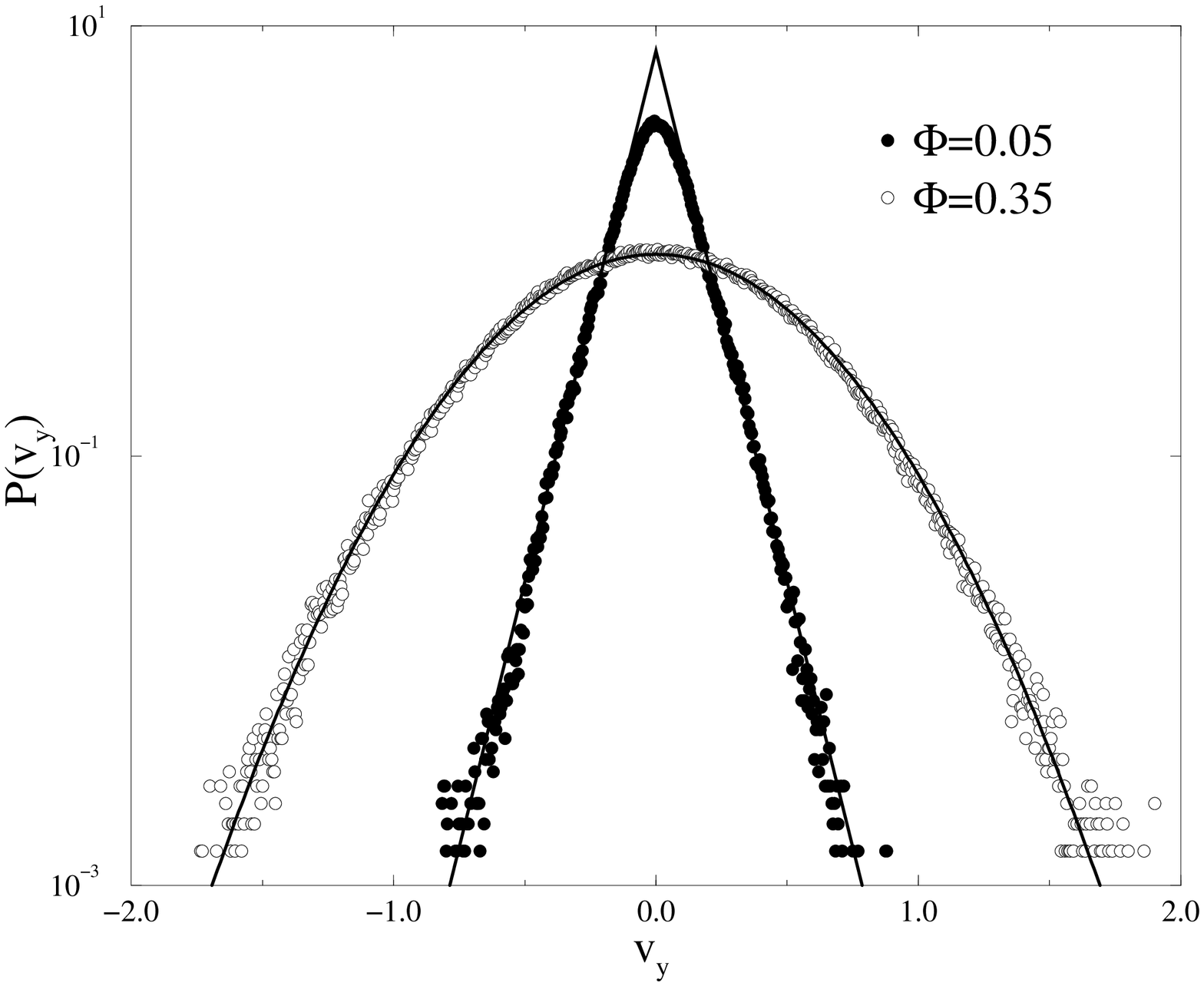}
\includegraphics[width=9.4cm,angle=-90]{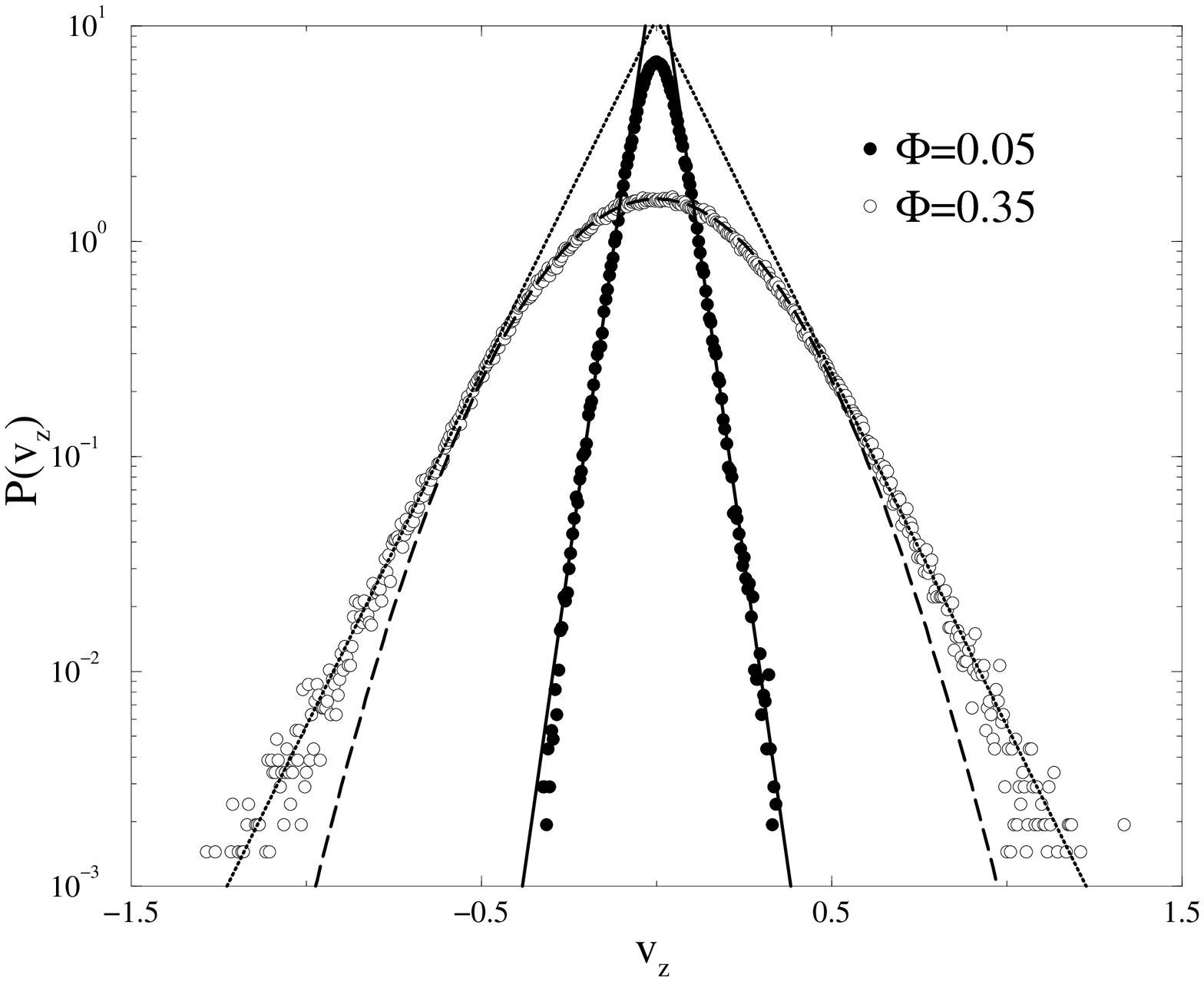}
\caption[pdfN]{Probability density function of the velocity fluctuations
in both transverse directions (a) $P(v_y)$ and (b) $P(v_z)$, 
for two different volume fractions $\phi = 0.05$ and $\phi =0.35$.
See \S\,\ref{fluctuations} for a discussion of the different 
distributions used to fit the numerical data (solid and dashed lines).
The simulations are for $N=64$, $N_c=100$, $F_0=1.0$ and $r_c=10^{-4}$.}
\label{pdf}
\end{figure}

Numerical simulations clearly suffer from two limitations,
which become of particular importance when computing
velocity fluctuations: the small number of particles in the
simulations and the necessary approximations to the hydrodynamic 
forces \cite[]{brady88}. However, we shall show that
the functional form of the pdf's undergoes a clear transition, from an exponential 
to a Gaussian distribution as the volume fraction is increased, which 
is not sensitive to the number of particles. 
Moreover, this feature could be measured experimentally 
thereby providing a useful test for the approximations involved in 
the hydrodynamic interactions which have been used in constructing
the Stokesian dynamics code.

In figure \ref{pdfN} we show the pdf of the velocity fluctuations in the
direction of the shear $P(v_y)$, obtained at two different
volume fractions, $\phi=0.05$ and $\phi=0.35$, and using different 
number of particles in the simulations. It can be seen that, as previously
stated, the distribution of velocities is insensitive to
the number of particles, at least as far as its functional form
is concerned.  

In figure \ref{pdf} we compare the pdf's at low and high volume
fractions in a log-linear plot. Clearly, both $P(v_y)$ and $P(v_z)$ become 
not only broader as the concentration increases but 
change noticeably in their shape. In figure \ref{pdf}a 
we show that, at low volume fractions ($\phi=0.05$), $P(v_y)$ is well described
by an exponential distribution, $P(v_y) \propto \exp(-11.4 \, |v_y|)$ but that,
at a large volume fraction, $\phi=0.35$,
a Gaussian distribution accurately fits the numerical results,
$P(v_y) \propto \exp(-2.41 \, v_y^2)$. 
A similar behaviour is observed in $P(v_z)$, as shown in figure
\ref{pdf}b in that, at low volume fractions, the pdf is well
approximated by an exponential distribution
$P(v_z) \propto \exp(-26.3 \, |v_z|)$, but at large concentrations
neither a Gaussian nor an exponential distribution accurately
fits the numerical results over the whole range of velocities.
However, we show that, for small velocity fluctuations, the distribution 
is Gaussian (dashed line; $P(v_z) \propto \exp(-7.76 \, v_z^2)$), 
whereas large fluctuations are better described by an 
exponential distribution (dotted line; $P(v_z) \propto \exp(-7.57 \, |v_z|)$).
A similar transition, but in the opposite direction, was 
observed in fludized suspensions of non-colloidal particles,
where the velocity fluctuation distributions vary
from Gaussian to exponential as the particle concentration increases
\cite[][]{rouyer99,rouyer00}.

We believe that this transition in the pdf
of the velocity fluctuations towards a Gaussian distribution, 
is due to the predominant role played by lubrication forces 
at the high concentrations. Recall that, given their short-range 
character, lubrication forces are essentially two-body interactions 
and are accounted for in a pairwise additive way in the simulation method. 
On the other hand, again in the simulation method, many-body long-range hydrodynamic 
interactions are accounted for by means of the far-field 
approximation\footnote{A detailed 
discussion of the approximations involved in the
Stokesian dynamics method can be found elsewhere
\cite[]{durlofsky87,brady88,brady88b,ichiki01}.}. 
Therefore, the random addition of lubrication forces, generated from
spheres in close proximity to one another, is the dominant contribution at high
concentrations, and would be expected to yield a Gaussian
distribution. Let us also note that the hydrodynamic forces
inhibit the relative motion of the particles as their approach
one another and therefore would prevent 
large fluctuations in the velocity from occurring, consistent with
the range in $v_z$ where a Gaussian distribution properly 
describes the corresponding pdf's. On the contrary, at low concentrations,
lubrication forces becomes negligible and therefore
long-range many-body interactions determine the fluctuations 
in velocity. In this case, our results show an exponential 
distribution. It is clear then, that a comparison with
experimental results would provide another test of the numerical method 
and of the approximations that are
involved in the simulations. 

An alternative view of the distinction between Gaussian and exponential
velocity distributions may be offered by analogy with turbulence. Specifically, in
boundary layer flows, one observes that the pdf of vorticity is
Gaussian at low Reynolds numbers, but develops exponential tails at the
very high Reynolds numbers characteristic of atmospheric flows \cite[][]{fan91}.
Similarly (perhaps), the pdf of temperature fluctuations in thermal
convection exhibits a transition from Gaussian to exponential as the
Rayleigh number increases \cite[][]{sano89,wu91}.  In both cases, the transition is gradual
rather than abrupt, and the increase of Reynolds or Rayleigh number is
accompanied by the development of organized large-scale coherent
structures in the flow \cite[][ p. 100]{frisch}.  Since one normally associates 
Gaussian pdf's with the addition of uncorrelated random variables, it is 
natural to relate the transition to the more slowly-decaying exponential pdf's 
to the appearance of correlated long-range structures.  In the case of
suspensions, inverse concentration is analogous to the Reynolds or
Rayleigh number in the sense that a large value of the parameter is
associated with long-ranged correlations, as we have seen in the 
discussion of screening in \S\,\ref{autocorrela}.  A common feature in these three
situations is that, at low values of the appropriate control parameter,
the motion is relatively uncorrelated and the fluctuations are 
Gaussian while, at high values, the motion is organized and the fluctuations are 
more persistent and exponentially distributed.

\section{Shear-induced self-diffusion at low concentrations}
\label{diffusion}

\begin{figure}
\centering
\includegraphics[width=10cm,angle=-90]{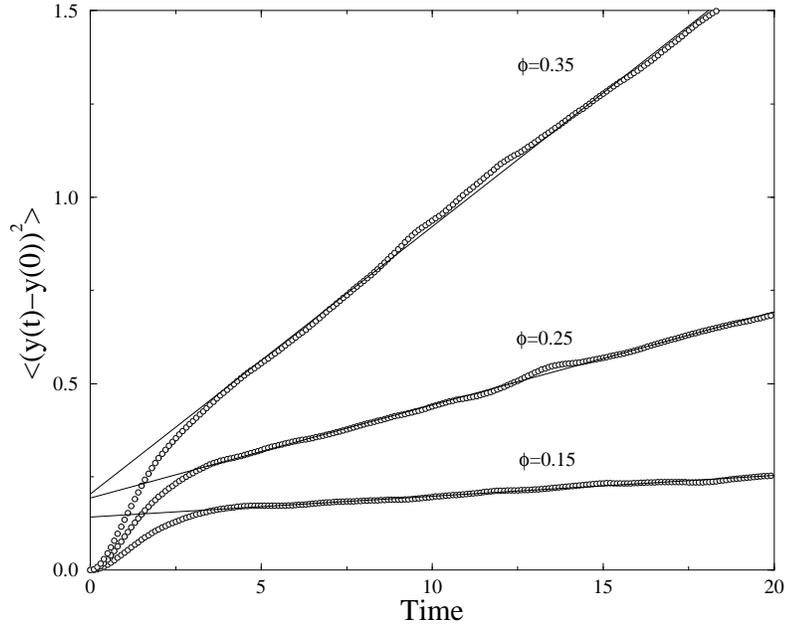}
\includegraphics[width=10cm,angle=-90]{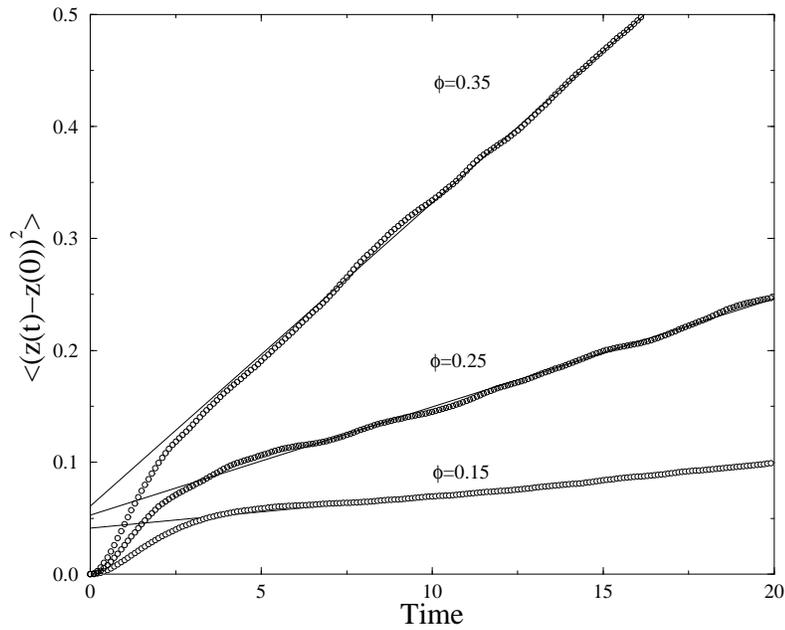}
\caption[]{Mean square displacement of a single particle,
averaged over all the $N=64$ particles in the suspension and
over $N_c=100$ different realizations, as a function of time for different
volume fractions. The magnitude of the interparticle force is $F_0=1$ 
and the characteristic range $r_c=10^{-4}$. }
\label{XvsT}
\end{figure}

In section \S\,\ref{chaos} we discussed how stochastic (diffusive-like) 
transport arises in the deterministic dynamics of sheared suspensions,
due to the chaotic evolution in phase space and,
in \S\,\ref{autocorrela}, we showed the presence of a stochastic motion
at the individual level of a single sphere by studying the
loss of correlation in the transverse particle velocities which leads to a random 
diffusive motion of single particles. Finally, in figure \ref{XvsT}, we show 
that the mean square displacement of a single particle becomes diffusive 
after a time approximately equal to the characteristic time after which 
the particle's velocity correlation is lost.

\begin{figure}
\centering
\includegraphics[width=10cm,angle=-90]{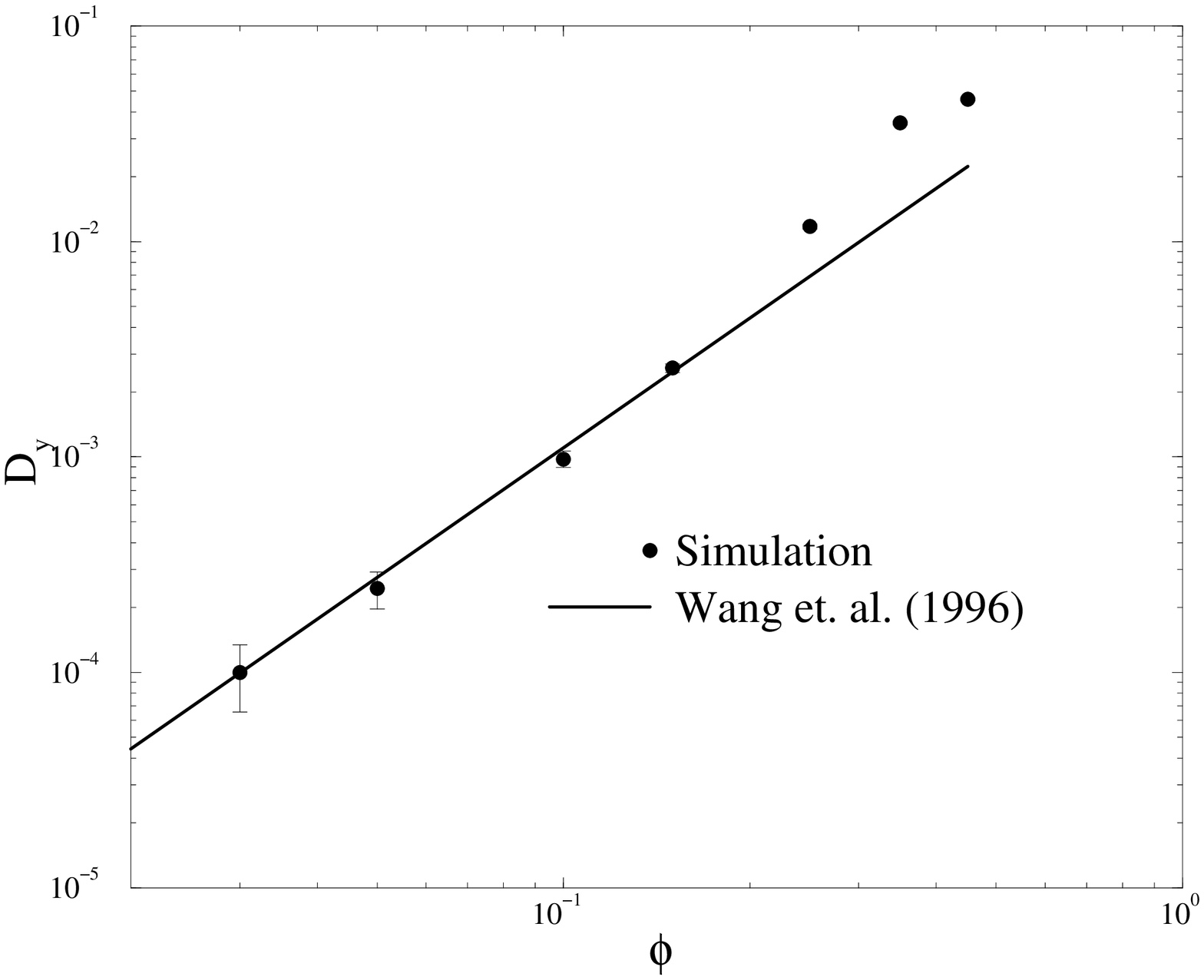}
\caption[dysmooth]{Dimensionless diffusion coefficient in the direction of the shear
$D_y$ as a function of the volume fraction. The solid line
correspond to the theoretical result calculated by \cite{wang96},
in the absence of non-hydrodynamic forces.
The simulations are for $N=64$, $N_c=100$, $F_0=1.0$ and $r_c=10^{-4}$.}
\label{dysmooth}
\end{figure}

We also discussed
the relation between the negative correlation in the transverse velocity 
fluctuations and the hydrodynamic interaction between 
a pair of non-Brownian spheres undergoing shear.
We also remarked that, in order to get a diffusive motion
from purely hydrodynamic interactions, collisions between
at least three particles are necessary while, on the contrary,
in the presence of non-hydrodynamic forces such 
as the interparticle repulsive force introduced in 
\S\,\ref{interparticle_force}, binary collisions alone lead to diffusive
motion. This fact gives rise to different scaling relations for
the diffusivity depending on the relative importance of the
interparticle force and the hydrodynamic forces. 
Let the diffusion coefficient in the pure
hydrodynamic limit be denoted by $D_h$, and the contribution to the diffusivity
in the presence of non-hydrodynamic forces by $D_{n-h}$.
As already mentioned in \S\,\ref{intro}, 
since the hydrodynamic contribution $D_h$ 
arises from collisions between three or more particles, and the rate at which 
a given sphere interacts with two other spheres is proportional 
to $\gamma \, \phi^2$ in the
limit of low volume fractions, $D_h$ should scale
as $\gamma \, \phi^2 \, a^2$ as $\phi \to 0$ while
$D_{n-h} \propto f(F_0,r_c) \; \gamma \, \phi \, a^2$, where $f(F_0,r_c)$ 
should be an increasing function of the interparticle force magnitude
$F_0$, and of the range $r_c$. Then, when interparticle forces are negligible,
the diffusion coefficient $D$ will be dominated by the
hydrodynamic contribution and should scale as $\gamma \, \phi^2 \, a^2$. 
On the other hand, at low enough concentrations or large
interparticle forces, binary collisions become dominant and a linear 
dependence on the volume fraction should be expected, i. e.
$D \sim D_{n-h} \propto \gamma \, \phi \, a^2$.

\begin{figure}
\centering
\includegraphics[width=10cm,angle=-90]{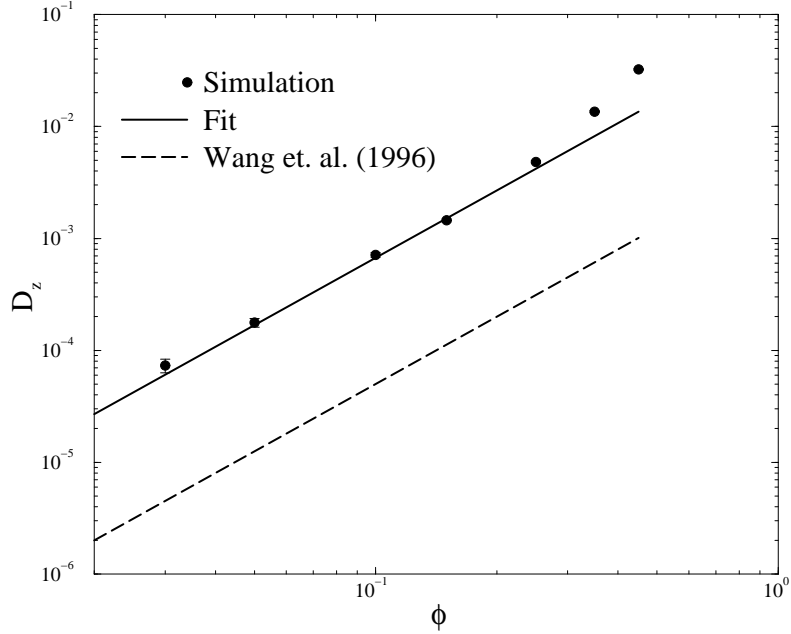}
\caption[dzsmooth]{Dimensionless diffusion coefficient 
$D_z$ as a function of the volume fraction.
The solid line corresponds to a fit $D_z=0.07 \, \phi^2$.
The dashed line corresponds to the result obtained by \cite{wang96},
$D_z=0.005 \, \phi^2$. 
The simulations are for $N=64$, $N_c=100$, $F_0=1.0$ and $r_c=10^{-4}$.}
\label{dzsmooth}
\end{figure}

Let us first investigate the case when the interparticle force is negligible.
In figure \ref{dysmooth} we present, in a log-log plot, 
$D_y$ the diffusion coefficient in the direction 
of the shear rendered dimensionless by $\gamma a^2$, 
as a function of the volume fraction where, in all cases, 
the diffusion coefficient is obtained from the
slope of the mean square displacement in the region of its linear growth,
as shown in figure \ref{XvsT}.
In the numerical simulations the characteristic range of the interparticle 
force was set to $r_c=10^{-4}$ and $F_0=1.0$.
It can be observed that at low volume fractions, a quadratic dependence
of $D_y$ on $\phi$ is obtained, as expected when the
non-hydrodynamic force is negligible. 
We also compare in figure \ref{dysmooth} the numerical 
results with the theoretical values obtained by \cite{wang96}, who 
evaluated the diffusion coefficient by 
computing the mean square displacement
of a test sphere over all possible encounters with two other spheres in the absence of
the non-hydrodynamic force and found that $D_y=0.11 \, \phi^2$, 
in excellent agreement with our fitted value $D_y=(0.11 \pm 0.02) \, \phi^2$.
However, it should be kept in mind that the diffusion coefficients
reported here are from simulations using only $64$ particles, 
a number which may not be large enough for an accurate comparison
with other vales of $D$ obtained theoretically or experimentally. 

In figure \ref{dzsmooth} we present the dependence of 
$D_z$ on $\phi$. In this case, a quadratic dependence is also found.
However, the theoretical values calculated by \cite{wang96}
are much smaller than our numerical results in that a fit to the numerical results 
gives $D_z=(0.07 \pm 0.007)\, \phi^2$, 
whereas \cite{wang96} found $D_z=0.005 \, \phi^2$.
However, let us note that the anisotropy found in our simulations
for the self-diffusion coefficient, $D_y/D_z \sim 1.5$, is in agreement with
the experimental results of \cite{phan99} recently confirmed
by \cite{breedveld2001b}.

\begin{figure}
\centering
\includegraphics[width=10cm,angle=-90]{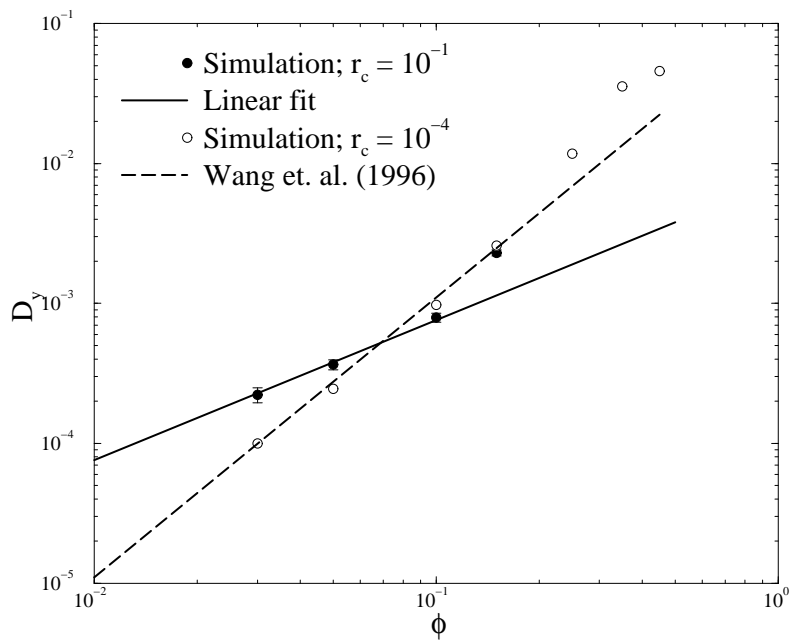}
\caption[dyrough]{Dimensionless diffusion coefficient in the plane of shear $D_y$ as
a function of the volume fraction $\phi$. Open circles corresponds
to short-ranged interparticle forces $r_c=10^{-4}$, while solid
circles correspond to a longer range $r_c=10^{-1}$. The solid
line is a best fit with a linear dependence on $\phi$: 
$D_y = (7.6\pm1.3)\times10^{-3} \, \phi$. 
The simulations are for $N=64$, $N_c=100$, and $F_0=1.0$.}
\label{dyrough}
\end{figure}

Let us now consider the case in which the interparticle force contributes 
significantly to the diffusion coefficient. 
To this end, we performed simulations with the
characteristic range of the interparticle force increased 1000 
times relative to its previous value $r_c=10^{-4}$.
The strength of the force was kept constant $F_0=1.0$.
In figure \ref{dyrough} we present the new numerical results 
for the diffusion coefficient in the direction of shear,
as a function of the volume fraction. For comparison we
also show the previous results.
It is clear that the diffusivity deviates from
the values found previously and becomes larger than before
at very low concentrations.
As $\phi \to 0$, a linear dependence on $\phi$ is found to 
accurately fit the numerical data, giving 
$D_y = (7.6\pm1.3)\times10^{-3} \, \phi$.

\begin{figure}
\centering
\includegraphics[width=10cm,angle=-90]{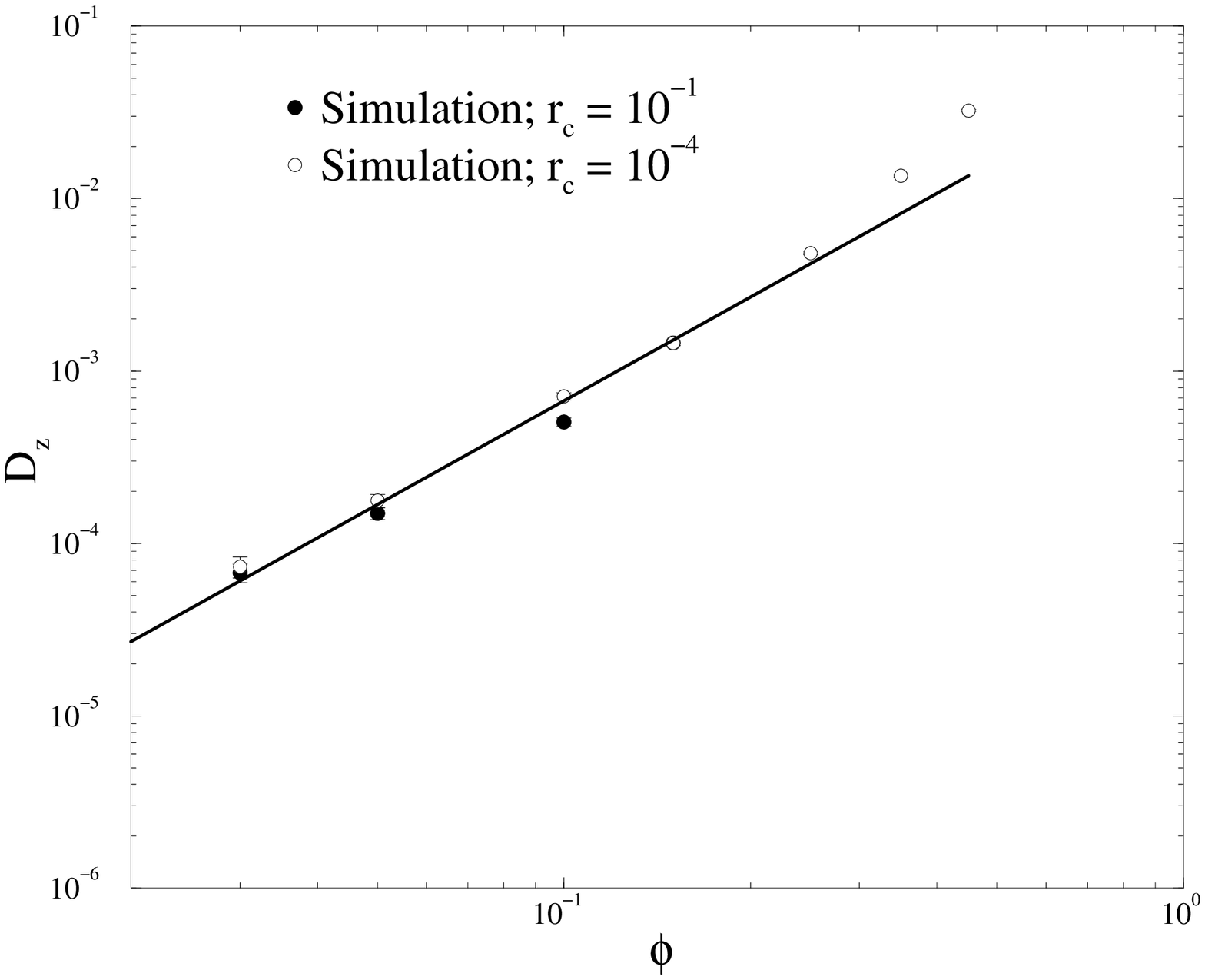}
\caption[dzrough]{Dimensionless diffusion coefficient perpendicular to the plane of 
shear $D_z$ as a function of the volume fraction $\phi$. 
Open and solid circles corresponds to 
different characteristic range of the interparticle force,
$r_c=10^{-4}$ and $r_c=10^{-1}$ respectively.
The solid line is the best fit with a quadratic dependence on $\phi$,
$D_z=(0.07 \pm 0.007)\, \phi^2$. 
The simulations are for $N=64$, $N_c=100$, and $F_0=1.0$.}
\label{dzrough}
\end{figure}

In figure \ref{dzrough} the diffusion coefficient in the vorticity plane
is compared for the same two values of the characteristic range,
$r_c=10^{-4}$ and $r_c=10^{-1}$, as a function of $\phi$.
In this case, the interparticle force is not strong enough and
the concentration not low enough for the linear regime to be observed.

\begin{figure}
\centering
\includegraphics[width=10cm,angle=-90]{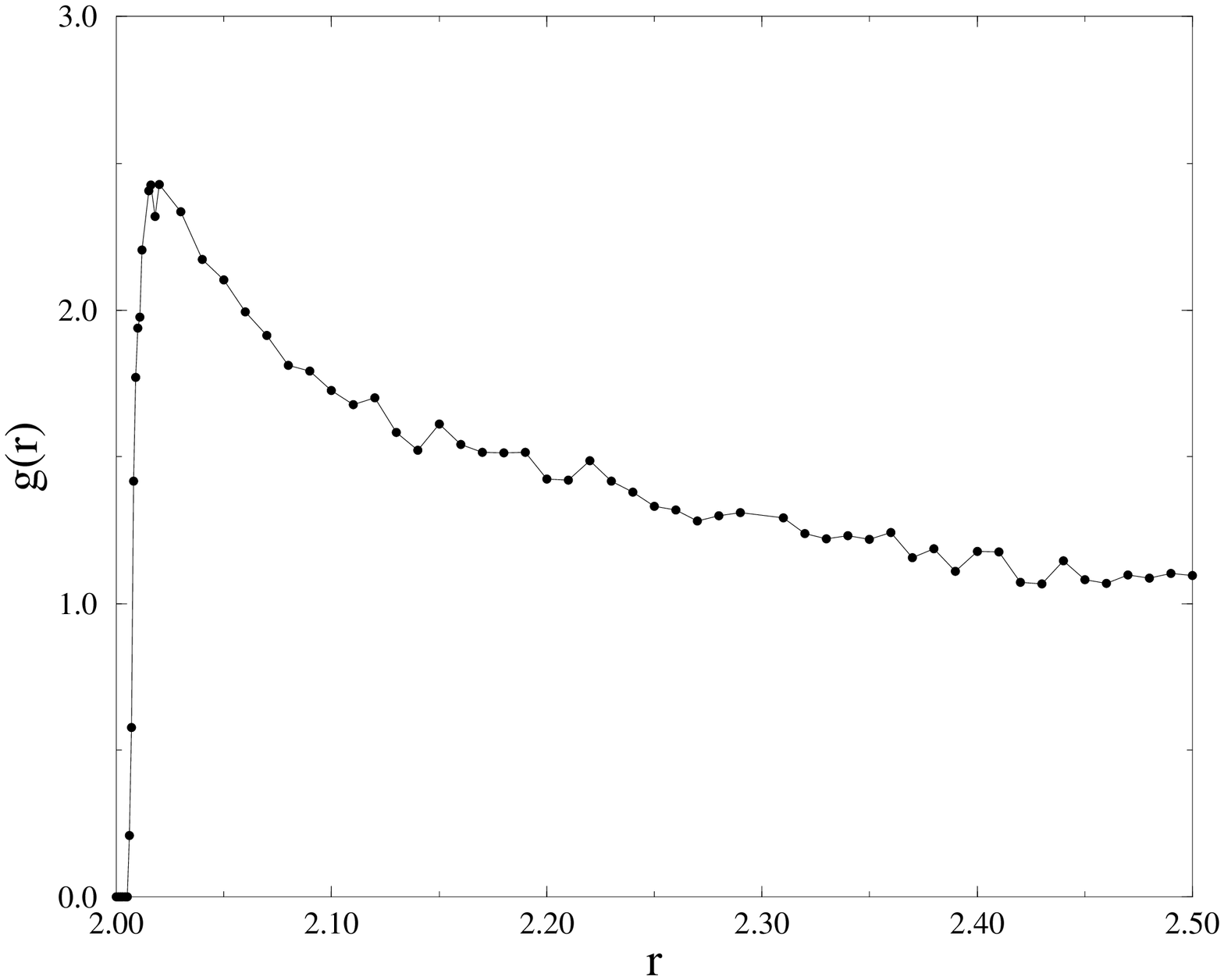}
\caption[gr_L1]{Pair distribution function $g(r)$. 
The characteristic range is $r_c=0.1$. 
$N=64$, $\phi=0.10$, $F_0=1.0$, and $N_c=100$.}
\label{gr_L1}
\end{figure}

It is important to note that the largest range of the interparticle force
$r_c=0.1$ is clearly too large to be considered as
representing residual Brownian forces or electrostatic repulsion
\cite[][]{foss00}. However, this force might resemble the effect 
of particle roughness. By means of numerical simulations, 
\cite{cunha96} and recently \cite{zarraga01},
showed that particle roughness lead to a diffusivity proportional to the
volume fraction. These authors modelled particle roughness by a
normal force which prevents the particles from coming closer than
a minimum dimensionless separation $2+\epsilon$ 
between the centres of the two spheres.
As a crude approximation, we might estimate the 
magnitude of the roughness represented by our
interparticle force, as its characteristic range $r_c$, i.e.  
$\epsilon \sim r_c \sim 0.1$ in our simulations.
For $\epsilon \sim 0.1$ \cite{cunha96} and \cite{zarraga01}
found that $D_y/\phi \sim 0.02$ , which is 
larger than our fitted value $D_y/\phi \sim (0.0076 \pm 0.0013)$ .
However, as shown in figure \ref{gr_L1}, in our simulations,
pairs of particles are allowed to come closer than $r_c$.This is  
because, at $r=r_c$ the interparticle force is not infinite as in the analysis
by \cite{cunha96} and \cite{zarraga01}. On the other hand,
if we consider the observed minimum distance between spheres as the 
effective roughness represented by our repulsive force then, 
$\epsilon \sim 0.01$ and from \cite{cunha96} $D_y/\phi \sim 0.006$ , 
which is now in fairly good agreement with our fitted value.
A linear behaviour $D_y \propto \phi$ was also found in experiments by \cite{biemfohr93}
and \cite{zarraga99}. However, the diffusion coefficient reported
in these experiments was substantially larger than that
obtained by means of numerical simulations ($D_y/\phi \sim 0.03$ as compared to
$D_y/\phi \sim 0.005$ from the numerical simulations for $\epsilon \sim 10^{-3}$.).

\section{Summary}
\label{summary}

The complex dynamics of homogeneous sheared suspensions of 
monodisperse, neutrally buoyant, non-Brownian spheres at 
zero Reynolds number was investigated by means of numerical
simulations using Stokesian dynamics. Starting from a
large number of independent initial configurations 
($N_c=100$), the evolution of typically $N=64$ spheres
undergoing simple linear shear was simulated during a
time $t \sim 100$, which was sufficiently long 
to allow us to study the dynamics of the system in 
steady state. In addition to the hydrodynamic interactions between spheres, 
the simulations included a short-ranged repulsive interparticle
force that qualitatively models the effects of surface roughness and
Brownian forces both of which play an important role when neighboring spheres
nearly touch one another. 

We began by recalling some of the well known effects of the non-hydrodynamic 
interaction on the microscopic structure of sheared suspensions
\cite[][]{bossis84,rampall97}. We showed that the location of the first peak of the pair distribution
function strongly depends on the range of the interparticle force $r_c$
in that, as $r_c$ is increased, the minimum separation between neighboring
spheres increases significantly. We also discussed how the angular distribution
of pairs depends on this force, showing that, as the range of the
interparticle force is increased, the pair distribution function
becomes increasingly more asymmetric, with fewer
pairs being oriented downstream than upstream. This asymmetry implies the loss of time
reversibility and therefore identifies the interparticle force as
a microscopic origin of irreversibility. 

The dynamics of sheared suspensions was shown to be chaotic over the whole
range of volume fractions investigated, $0.01 < \phi < 0.35$, with the 
largest Lyapunov exponent increasing linearly with $\phi$. 
The existence of chaos in the dynamics of the suspensions provides 
an explanation for the loss of correlation in the particle motions 
leading to diffusive behaviour
at long times. In fact, we showed that the system loses memory
of its initial state exponentially, and that the asymptotic separation 
distance between two initially close trajectories in the velocity-space
of the two transverse velocity components can be accurately estimated 
assuming that the two systems are 
completely uncorrelated. 

At the level of individual smooth spheres, we also demonstrate the loss of memory of
their instantaneous velocity fluctuations via the transverse velocity autocorrelation 
function. For high concentration, $\phi \sim 0.40$, a screening-type mechanism is 
observed and the correlations in particle velocities are lost after a 
short period of time of $O(1)$. On the other hand, at the lower concentrations,
the transverse particle motions remain, on average, correlated for a longer 
period of time. We found furthermore that, 
for both the transverse components of the velocity, 
the autocorrelation function becomes negative at intermediate times 
($t\sim 2$). We proposed an explanation for 
this effect based in the dynamics of two isolated spheres 
undergoing simple shear, according to which, since the purely hydrodynamic 
interaction between 
two spheres does not lead to any net lateral displacement, it is a source of
negative correlation in the velocity fluctuations during binary collisions. 
This explanation is consistent with the fact that the region within
which the velocity autocorrelation function is negative 
continues to enlarge with decreasing $\phi$ with the whole 
autocorrelation function appearing to converge to an asymptotic distribution 
dominated by binary interactions between spheres. 
We mentioned that the integral of this asymptotic distribution must
vanish on account of its being proportional to the net displacement experienced
by a pair of spheres, and showed that, in fact, the numerical value of the integral
of the autocorrelation function steadily decreases as $\phi \to 0$.
An important consequence is that the leading contribution to the 
diffusivity of the particles as $\phi \to 0$ comes from an
asymptotically negligible contribution to the autocorrelation function.
Therefore, an estimate of the diffusion coefficient,
based on the velocity autocorrelation function,
will be highly inaccurate at least for low volume fractions.

At this microscopic level we also computed the velocity probability distribution 
function in both lateral directions and observed a transition from an exponential 
to a Gaussian distribution
as the volume fraction is increased. We proposed that this transition
is due to the dominant role, at high concentrations, of the lubrication forces, 
which are essentially two-body random interactions and therefore 
would be expected to give 
rise to a Gaussian distribution. Unfortunately, there are no 
experimental measurements thus far of the probability distribution
function of the velocity fluctuations.
 
Finally, we investigated the scaling of the diffusion coefficient $D$
at low concentrations as the range of the interparticle force 
is varied, by setting $D$ equal to one half the slope
of the mean square displacement in the region of its linear growth. 
For a very small range, $r_c=10^{-4}$, we showed that
the effect of the interparticle force is negligible and
$D$ was found to scale as $\gamma \phi^2 a^2$ for $0.03 < \phi <0.15$,
both along and normal to the plane of shear. This results corresponds to a diffusive motion 
arising from collisions between three spheres simultaneously and,
in the case of $D_y$, is in very good agreement with the values obtained
by \cite{wang96}, who evaluated the diffusion coefficient
by a completely different procedure, viz. by 
computing the mean square displacement
of a test sphere over all possible encounters with two other spheres.
On the other hand, for a much larger range of the interparticle force,
$r_c=10^{-1}$, a linear dependence of the diffusion coefficient on $\phi$
is observed as $\phi \to 0$, which implies that a
significant contribution to the diffusion coefficient emanated from 
encounters between two particles. This is consistent with the
notion that the interparticle force qualitatively mimics the effects 
of surface roughness and other non-hydrodynamic forces, and  
reproduces the scaling behaviour found in the theoretical analysis
by \cite{cunha96} and in the experimental work by \cite{zarraga99}.
However, the numerical simulations still fail to reproduce the experimental 
values of the diffusivity reported by \cite{zarraga99} by an order of magnitude.

\begin{acknowledgements}

We thank Professor K. Sreenivasan for discussions on the common features 
between our data and turbulence, and Dr. J.~R.~Melrose and Professor J.~F.~Brady
for the use of their simulation codes. G. D. thanks M. Tirumkudulu and I. Baryshev
for their helpful comments. A.A. and G.D. were partially supported by the National 
Science Foundation under Grant CTS-9711442 and by the Engineering Research Program, 
Office of Basic Energy and Sciences, U.S. Department of Energy under Grant 
DE-FG02-90ER14139; J.K was partially supported by the Geosciences Research Program, 
Office of Basic Energy and Sciences, U.S. Department of Energy;
G. D. was partially supported by CONICET Argentina and The
University of Buenos Aires.
Computational facilities
were provided by the National Energy Resources Scientific
Computer Center.
\end{acknowledgements}

\clearpage
\newpage

\bibliography{../../../LATEX/biblio/article,../../../LATEX/biblio/book}

\begin{thebibliography}{50}
\expandafter\ifx\csname natexlab\endcsname\relax\def\natexlab#1{#1}\fi

\bibitem[Acrivos {\em et~al.\/}(1992)Acrivos, Batchelor, Hinch, Hoch \&
  Mauri]{acrivos92}
{\sc Acrivos, A., Batchelor, G.~K., Hinch, E.~J., Hoch, D.~L. \& Mauri, R.}
  1992 Longitudinal shear-induced diffusion of spheres in a dilute suspension.
  {\em J. Fluid Mech.\/} {\bf 240}, 651--657.

\bibitem[Alder \& Wainwright(1958)]{alder58}
{\sc Alder, B.~J. \& Wainwright, T.} 1958 Molecular dynamiocs by electronic
  computers. In {\em Proceedings of the International Symposium on Transport
  Processes in Statistical Mechanics\/} (ed. I.~Prigogine), pp. 97--131. New
  York: Interscience publishers.

\bibitem[Alder \& Wainwright(1967)]{alder67}
{\sc Alder, B.~J. \& Wainwright, T.} 1967 Velocity autocorrelations for hard
  spheres. {\em Phys. Rev. Lett.\/} {\bf 18}~(23), 988--990.

\bibitem[Alder \& Wainwright(1970)]{alder70}
{\sc Alder, B.~J. \& Wainwright, T.} 1970 Decay of the velocity autocorrelation
  function. {\em Phys. Rev. A\/} {\bf 1}~(1), 18--21.

\bibitem[Allen \& Tildesley(1987)]{allen}
{\sc Allen, M.~P. \& Tildesley, D.~J.} 1987 {\em Computer Simulation of
  Liquids\/}. Oxford: Clarendon Press.

\bibitem[Baker \& Gollub(1990)]{baker}
{\sc Baker, G.~L. \& Gollub, J.~P.} 1990 {\em Chaotic Dynamics: an
  introduction\/}. Cambridge: Cambridge University Press.

\bibitem[Benettin {\em et~al.\/}(1976)Benettin, Galgani \&
  Strelcyn]{benettin76}
{\sc Benettin, G., Galgani, L. \& Strelcyn, J.-M.} 1976 Kolmogorov entropy and
  numerical experiments. {\em Phys. Rev. A\/} {\bf 14}~(6), 2338--2345.

\bibitem[Biemfohr {\em et~al.\/}(1993)Biemfohr, Looby \& Leighton]{biemfohr93}
{\sc Biemfohr, S., Looby, T. \& Leighton, D.~T.} 1993 Measurement of the
  shear-induced coefficient of self-diffusion in dilute suspensions. In {\em
  Proc. DOE/NSF Workshop on Flow of Particles and Fluids\/}. Ithaca, New York.

\bibitem[Bossis \& Brady(1984)]{bossis84}
{\sc Bossis, G. \& Brady, J.~F.} 1984 Dynamic simulation of sheared
  suspensions. i. general method. {\em J. Chem. Phys.\/} {\bf 80}~(10),
  5141--5161.

\bibitem[Brady \& Bossis(1984)]{brady84}
{\sc Brady, J.~F. \& Bossis, G.} 1984 Dynamic simulation of sheared
  suspensions. i. general method. {\em J. Chem. Phys.\/} {\bf 80}~(10),
  5141--5152.

\bibitem[Brady \& Bossis(1985)]{brady85}
{\sc Brady, J.~F. \& Bossis, G.} 1985 The rheology of concentrated suspensions
  of spheres in simple shear flow by numerical simulation. {\em J. Fluid
  Mech.\/} {\bf 155}, 105--129.

\bibitem[Brady \& Bossis(1988)]{brady88}
{\sc Brady, J.~F. \& Bossis, G.} 1988 Stokesian dynamics. {\em Annu. Rev. Fluid
  Mech.\/} {\bf 20}, 111--140.

\bibitem[Brady \& Morris(1997)]{brady97}
{\sc Brady, J.~F. \& Morris, J.~F.} 1997 Microstructure of strongly sheared
  suspensions and its impact on rheology and diffusion. {\em J. Fluid Mech.\/}
  {\bf 348}, 103--139.

\bibitem[Brady {\em et~al.\/}(1988)Brady, Phillips, Lester \& Bossis]{brady88b}
{\sc Brady, J.~F., Phillips, R.~J., Lester, J.~C. \& Bossis, G.} 1988 Dynamic
  simulation of hydrodynamically interacting suspensions. {\em J. Fluid
  Mech.\/} {\bf 195}, 257--280.

\bibitem[Breedveld {\em et~al.\/}(2001{\natexlab{{\em a\/}}})Breedveld, {van
  den Ende}, Bosscher, Lomgschaap \& Mellena]{breedveld2001b}
{\sc Breedveld, V., {van den Ende}, D., Bosscher, M.~B., Lomgschaap, J. J.~J.
  \& Mellena, J.} 2001{\natexlab{{\em a\/}}} Measuring shear-induced
  self-diffusion in a counterrotating geometry. {\em Phys. Rev. E\/} {\bf 63},
  021403.

\bibitem[Breedveld {\em et~al.\/}(2001{\natexlab{{\em b\/}}})Breedveld, {van
  den Ende}, Lomgschaap \& Mellena]{breedveld2001a}
{\sc Breedveld, V., {van den Ende}, D., Lomgschaap, J. J.~J. \& Mellena, J.}
  2001{\natexlab{{\em b\/}}} Shear-induced diffusion and rheology of
  noncolloidal suspensions: Timescales and particle displacements. {\em J.
  Chem. Phys.\/} {\bf 114}, 5923--5936.

\bibitem[Breedveld {\em et~al.\/}(1998)Breedveld, {van den Ende}, Tripathi \&
  Acrivos]{breedveld98}
{\sc Breedveld, V., {van den Ende}, D., Tripathi, A. \& Acrivos, A.} 1998 The
  measurement of the shear-induced particle and fluid tracer diffusivities in
  concentrated suspensions by a novel method. {\em J. Fluid Mech.\/} {\bf 375},
  297--318.

\bibitem[Brenner \& Mucha(2001)]{brenner01}
{\sc Brenner, M.~P. \& Mucha, P.~J.} 2001 That sinking feeling. {\em Nature\/}
  {\bf 409}, 568--570.

\bibitem[Brinkman(1947)]{brinkman47}
{\sc Brinkman, H.~C.} 1947 A calculation of the viscous force exerted by a
  flowing fluid on a dense swarm of particles. {\em Appl. Sci. Res.\/} {\bf
  A}~(1), 27--34.

\bibitem[{da Cunha} \& Hinch(1996)]{cunha96}
{\sc {da Cunha}, F.~R. \& Hinch, E.~J.} 1996 Shear-induced dispersion in a
  dilute suspension of rough spheres. {\em J. Fluid Mech.\/} {\bf 309},
  211--223.

\bibitem[Dorfman(1998)]{dorfman98}
{\sc Dorfman, J.~R.} 1998 Deterministic chaos and the foundations of the
  kinetic theory of gases. {\em Phys. Rep.\/} {\bf 301}, 151--185.

\bibitem[Dratler \& Schowalter(1996)]{dratler96}
{\sc Dratler, D.~I. \& Schowalter, W.~R.} 1996 Dynamic simulation of
  suspensions of non-brownian hard spheres. {\em J. Fluid Mech.\/} {\bf 325},
  53--77.

\bibitem[Durlofsky \& Brady(1987)]{durlofsky87b}
{\sc Durlofsky, L. \& Brady, J.~F.} 1987 Analysis of the brinkman equation as a
  model for flow in porous media. {\em Phys. Fluids\/} {\bf 30}~(11),
  3329--3341.

\bibitem[Durlofsky {\em et~al.\/}(1987)Durlofsky, Brady \& Bossis]{durlofsky87}
{\sc Durlofsky, L., Brady, J.~F. \& Bossis, G.} 1987 Dynamic simulation of
  hydrodynamically interacting particles. {\em J. Fluid Mech.\/} {\bf 180},
  21--49.

\bibitem[Eckmann \& Ruelle(1985)]{ruelle85}
{\sc Eckmann, J.~P. \& Ruelle, D.} 1985 Ergodic theory of chaos and strange
  attractors. {\em Rev. Mod. Phys.\/} {\bf 57}~(3), 617--656.

\bibitem[Eckstein {\em et~al.\/}(1977)Eckstein, Bailey \& Shapiro]{eckstein77}
{\sc Eckstein, E.~C., Bailey, D.~G. \& Shapiro, A.~H.} 1977 Self-diffusion of
  particles in shear flow of a suspension. {\em J. Fluid Mech.\/} {\bf 79}~(1),
  191--208.

\bibitem[Fan(1991)]{fan91}
{\sc Fan, M.~S.} 1991 Features of vorticity in fully turbulent flows. PhD
  thesis, Yale University.

\bibitem[Foss \& Brady(2000)]{foss00}
{\sc Foss, D.~R. \& Brady, J.~F.} 2000 Structure, diffusion and rheology of
  brownian suspensions by stokesian dynamics simulation. {\em J. Fluid Mech.\/}
  {\bf 407}, 167--200.

\bibitem[Frisch(1995)]{frisch}
{\sc Frisch, U.} 1995 {\em Turbulence: the legacy of A. N. Kolmogorov\/}.
  Cambridge: Cambridge University Press.

\bibitem[Gadala-Maria \& Acrivos(1980)]{gadala80}
{\sc Gadala-Maria, F. \& Acrivos, A.} 1980 Sheared-induced structure in a
  concentrated suspension of solid spheres. {\em J. Rheology\/} {\bf 24},
  799--814.

\bibitem[Gaspard(1998)]{gaspard}
{\sc Gaspard, P.} 1998 {\em Chaos, scattering and statistical mechanics\/}. New
  York: Cambridge University Press.

\bibitem[Ichiki \& Brady(2001)]{ichiki01}
{\sc Ichiki, K. \& Brady, J.~F.} 2001 Many-body effects and matrix inversion in
  low-reynolds-number hydrodynamics. {\em Phys. Fluids\/} {\bf 13}~(1),
  350--353.

\bibitem[van Kampen(1987)]{vankampen}
{\sc van Kampen, N.~G.} 1987 {\em Stochastic Processes in Physics and
  Chemistry\/}. Amsterdam: North-Holland.

\bibitem[Leal(1992)]{leal}
{\sc Leal, L.~G.} 1992 {\em Laminar Flow and Convective Transport Processes\/}.
  Boston: Butterworth-Heinemann.

\bibitem[Leighton \& Acrivos(1987)]{leighton87a}
{\sc Leighton, D. \& Acrivos, A.} 1987 Measurement of shear-induced
  self-diffusion in concentrated suspensions of spheres. {\em J. Fluid Mech.\/}
  {\bf 177}, 109--131.

\bibitem[Marchioro \& Acrivos(2001)]{marchioro2001}
{\sc Marchioro, M. \& Acrivos, A.} 2001 Shear-induced particle diffusivities
  from numerical simulations. {\em J. Fluid Mech.\/} To appear.

\bibitem[Nicolai {\em et~al.\/}(1995)Nicolai, Herzhaft, Hinch, Onger \&
  Guazzelli]{nicolai95}
{\sc Nicolai, H., Herzhaft, B., Hinch, E.~J., Onger, L. \& Guazzelli, E.} 1995
  Particle velocity fluctuations and hydrodynamic self-diffusion of sedimenting
  non-brownian spheres. {\em Phys. Fluids\/} {\bf 7}~(1), 12--23.

\bibitem[Parisi \& Gadala-Maria(1987)]{parisi87}
{\sc Parisi, F. \& Gadala-Maria, F.} 1987 Fore-and-aft asymmetry in a
  concentrated suspension of solid spheres. {\em J. Rheol.\/} {\bf 31}~(8),
  725--732.

\bibitem[Phan \& Leighton(1999)]{phan99}
{\sc Phan, S.~E. \& Leighton, D.~T.} 1999 Measurement of the shear-induced
  tracer diffusivity in concentrated suspensions. {\em J. Fluid Mech.\/}
  Submitted.

\bibitem[Rahman(1964)]{rahman64}
{\sc Rahman, A.} 1964 Correlations in the motion of atoms in liquid argon. {\em
  Phys. Rev.\/} {\bf 136}~(2A), 405--411.

\bibitem[Rampall {\em et~al.\/}(1997)Rampall, Smart \& Leighton]{rampall97}
{\sc Rampall, I., Smart, J.~R. \& Leighton, D.~T.} 1997 The influence of
  surface roughness on the particle-pair distribution function of dilute
  suspensions of non-colloidal spheres in simple shear flow. {\em J. Fluid
  Mech.\/} {\bf 339}, 1--24.

\bibitem[Rouyer {\em et~al.\/}(2000)Rouyer, Lhuillier, Martin \&
  Salin]{rouyer00}
{\sc Rouyer, F., Lhuillier, D., Martin, J. \& Salin, D.} 2000 Structure,
  density, and velocity fluctuations in quasi-two-dimensional non-brownian
  suspensions of spheres. {\em Phys. Fluids\/} {\bf 12}~(5), 958--963.

\bibitem[Rouyer {\em et~al.\/}(1999)Rouyer, Martin \& Salin]{rouyer99}
{\sc Rouyer, F., Martin, J. \& Salin, D.} 1999 Non-gaussian dynamics in
  quasi-2d noncolloidal suspension. {\em Phys. Rev. Lett.\/} {\bf 83}~(5),
  1058--1061.

\bibitem[Sano {\em et~al.\/}(1989)Sano, Wu \& Libchaber]{sano89}
{\sc Sano, M., Wu, X.~Z. \& Libchaber, A.} 1989 Turbulence in helium-gas free
  convection. {\em Phys. Rev. A\/} {\bf 40}~(11), 6421--6430.

\bibitem[Schuster(1989)]{schuster}
{\sc Schuster, H.~G.} 1989 {\em Deterministic chaos\/}. Weinheim: VCH.

\bibitem[Wang {\em et~al.\/}(1996)Wang, Mauri \& Acrivos]{wang96}
{\sc Wang, Y., Mauri, R. \& Acrivos, A.} 1996 The transverse shear-induced
  liquid and particle tracer diffusivities of a dilute suspension of spheres
  undergoing a simple shear flow. {\em J. Fluid Mech.\/} {\bf 327}, 255--272.

\bibitem[Wang {\em et~al.\/}(1998)Wang, Mauri \& Acrivos]{wang98}
{\sc Wang, Y., Mauri, R. \& Acrivos, A.} 1998 Transverse shear-induced gradient
  diffusion in a dilute suspension of spheres. {\em J. Fluid Mech.\/} {\bf
  357}, 279--287.

\bibitem[Wu(1991)]{wu91}
{\sc Wu, X.~Z.} 1991 Along a road to developed turbulence: free thermal
  convection in low temperature helium gas. PhD thesis, Chicago University.

\bibitem[Zarraga \& Leighton(1999)]{zarraga99}
{\sc Zarraga, I.~E. \& Leighton, D.~T.} 1999 Anomalous diffusion in a dilute
  suspension of noncolloidal spheres. Unpublished.

\bibitem[Zarraga \& Leighton(2001)]{zarraga01}
{\sc Zarraga, I.~E. \& Leighton, D.~T.} 2001 Normal stress and diffusion in a
  dilute suspension of hard spheres undergoing simple shear. {\em Phys.
  Fluids\/} {\bf 13}~(3), 565--577.

\end{thebibliography}
\bibliographystyle{jfm}

\end{document}